\documentclass[structabstract]{aa}  
\usepackage{txfonts} 
\usepackage{mathabx}
\usepackage{graphicx}
\usepackage{natbib} 
\bibpunct{(}{)}{;}{a}{}{,}
\usepackage{color}
\usepackage{amsmath}
\usepackage{graphicx,picture,calc}
\usepackage{lipsum}
\usepackage{footnote}
\usepackage{threeparttable}

\begin{document}

\title{Exoplanetary atmospheric sodium revealed by orbital motion}

\subtitle{Narrow-band transmission spectroscopy of \mbox{HD~189733b} with UVES}

\author{S. Khalafinejad\inst{1}, C. von Essen\inst{2}, H. J. Hoeijmakers\inst{3}, G. Zhou\inst{4}, T. Klocov\'a\inst{1}, J.H.M.M. Schmitt\inst{1}, S. Dreizler\inst{5}, M. Lopez-Morales\inst{4}, T.-O. Husser\inst{5}, T.O.B. Schmidt\inst{1}, R. Collet\inst{2}
}
\institute{Hamburg Observatory, Hamburg University, Gojenbergsweg 112, 21029 Hamburg, Germany \and 
Stellar Astrophysics Centre, Department of Physics and Astronomy, Aarhus University, Ny Munkegade 120, DK-8000 Aarhus C, Denmark \and Leiden Observatory, Leiden University, P.O. Box 9513, 2300 RA Leiden, The Netherlands \and 
Harvard-Smithsonian Center for Astrophysics, 60 Garden Street, Cambridge, MA 01238, USA \and Institute for Astrophysics, G\"ottingen University, Friedrich-Hund-Platz 1, 37077 G\"ottingen, Germany
}

%Context.       
\abstract{During primary transits, the spectral signatures of 
  exoplanet atmospheres can be measured using transmission spectroscopy. We can obtain information on the upper
  atmosphere of these planets by investigating the exoplanets' excess sodium  absorption in
the optical region. However, a number of factors can affect the observed sodium absorption signature. We present a detailed model correcting for systematic biases to yield an accurate depth for the sodium absorption in \mbox{HD 189733b}.
  }
%Aim.
{The goal of this work is to accurately measure the atomspheric sodium absorption light curve in HD 189733b, correcting for the effects of stellar differential limb-darkening, stellar activity, and a ``bump'' caused by the changing radial velocity of the exoplanet. In fact, owing to the high
  cadence and quality of our data, it is the first time that the last
  feature can be detected even by visual inspection.}
%Method.
{We use 244 high-resolution optical spectra taken by the UVES
  instrument mounted at the VLT. Our observations cover a
  full transit of \mbox{HD~189733b}, with a cadence of 45 seconds. To
  probe the transmission spectrum of sodium we produce excess
  light curves integrating the stellar flux in passbands of \mbox{1$\AA$}, \mbox{1.5 $\AA$}, and \mbox{3 $\AA$} inside the core of
  each sodium D-line. We model the effects of external sources on
  the excess  light curves, which correspond to an observed stellar
  flare beginning close to mid-transit time
  and the wavelength dependent limb-darkening effects. In addition, by
  characterizing the effect of the changing radial velocity and
  Doppler shifts of the planetary sodium lines inside the stellar
  sodium lines, we estimate the depth and width of the exoplanetary
  sodium feature.}
%Results.
{We estimate the shape of the planetary sodium line by a Gaussian profile with an equivalent width of  \mbox{$\sim 0.0023 \pm 0.0010$ $\AA$}, thereby confirming the presence of sodium in the atmosphere of HD 189733b with excess absorption levels of  \mbox{0.72 $\pm$
    0.25 \%}, \mbox{0.34 $\pm$ 0.11 \%}, and \mbox{0.20 $\pm$
    0.06 \%} for the integration bands of \mbox{1 $\AA$}, \mbox{1.5
    $\AA$}, and \mbox{3 $\AA$}, respectively. Using the equivalent width of the planetary sodium line, we produce a first order estimate
  of the number density of sodium in the exoplanet atmosphere.
  }
{}

\date{\today}

\maketitle

%---------------------------------------------------------------------------------------------
%                               INTRODUCTION
%---------------------------------------------------------------------------------------------
\section{Introduction}
\label{sec:introduction}

The first transiting hot Jupiter, \mbox{HD~209458b}, was discovered
more than a decade ago \citep{Charbonneau2000}. Since then a few
thousand transiting exoplanets have been detected with various transit
surveys from ground-based facilities
\citep[e.g.,][]{Hartman2004,Pollacco2006}, and from space
\citep[e.g.,][]{Borucki2010,Koch2010}. This has provided a fertile
ground for studying the exoplanet population from a global point of view
\citep{Sing2016}, and has opened a window to a deeper characterization
of planetary systems.

Transiting systems allow the study of their atmospheres via reflection
and transmission spectroscopy \citep[see the review paper by][and
  references therein]{Burrows2014}. During primary transit, the
exoplanet blocks part of the stellar light. If the planet has an
atmosphere, an additional fraction of the stellar light can be
absorbed by the materials present in its atmosphere. Since this
absorption takes place at discrete wavelengths, the planet size and
the transit light-curve depth will appear larger or smaller, depending
on the wavelength of the  observation. As a result, the variations in the
transit depth as a function of wavelength can reveal the presence of
different atomic and molecular species
\citep[e.g.,][]{Charbonneau2002,Tinetti2007,Vidal-Madjar2011,Crossfield2011,
  Desert2009,Hoeijmakers2015,Nikolov2015}, clouds
\citep{Gibson2013,Kreidberg2014}, and hazes \citep{Pont2013}.

Most of the spectro-photometric measurements of exoplanet atmospheres
have been performed through space-based observations
\citep[e.g.,][]{Deming2013,Sing2016}. Although these instruments have
superior stability and sensitivity, their low spectral resolution can
only constrain some atmospheric models. So far, high-resolution
spectroscopy has only been performed using ground-based facilities,
and despite the added complications posed by the Earth's atmosphere,
ground-based high-resolution studies have added valuable contributions
to our understanding of exoplanetary atmospheres. For example, at
resolutions of \mbox{$R\sim10^5$}, the absorption lines of individual
chemical species can be spectroscopically resolved
\citep{Birkby2013,Rodler2012,deKok2013,Lockwood2014,Brogi2012}, as can the
orbital motion, diurnal rotation of the planet, and exo-atmospheric
wind speed \citep{Snellen2010,Snellen2015,Louden2015,Brogi2012}. Even
large-scale dynamics in the upper atmosphere have already been
detected \citep[see, e.g.,][]{Kulow2014,Ehrenreich2015}.

Studies of the optical transmission spectra (mainly performed by the
Hubble Space Telescope, HST) have revealed that many hot Jupiters have
featureless spectra with strong Rayleigh scattering slopes toward
short wavelengths
\citep{Pont2008,Sing2011,Huitson2012,Nikolov2015,Sing2016}. This has
been attributed to the presence of high-altitude clouds and haze
layers that effectively obscure absorption features by making the
upper atmosphere opaque. In these cases, only species that are
present in the uppermost layers of the atmosphere can be
observed. Indeed, models indicate that only a handful of such species
are expected to be present in the optical region \citep[see,
  e.g.,][]{Fortney2010,Seager2000,Brown2001},  in consequence making
neutral sodium one of the most intensively studied alkali metals  to date.

Exoplanetary sodium absorption was first detected by \citet{Charbonneau2002} in the
atmosphere of \mbox{HD~209458b} by  analyzing spectro-photometric data
obtained by the HST during transit. Then, \cite{Redfield2008} claimed
the first ground-based detection of sodium in \mbox{HD~189733b} and in
the same year, \cite{Snellen2008} reported the ground-based detection of sodium in
\mbox{HD~209458b}. Altogether, sodium has been detected in a handful
of hot Jupiter atmospheres: \mbox{HD~209458b}
\citep{Charbonneau2002,Snellen2008}, \mbox{HD~189733b}
\citep{Redfield2008,Jensen2011,Huitson2012}, \mbox{WASP-17b}
\citep{Wood2011,Zhou2012}, and \mbox{XO-2b} \citep{Sing2012} are
classical examples. Recently, \cite{Wyttenbach2015} reported the
ground-based detection of sodium in \mbox{HD~189733b} using
observations collected during three transits with the HARPS
spectrograph. \cite{Cauley2016} also analyzed the pre-transit and in-transit phases at different atomic absorption lines, consisting of sodium.

In this paper, similar to \cite{Redfield2008} and
\cite{Wyttenbach2015}, we present the results of our efforts in the
detection of atmospheric sodium in the transmission spectrum of the
hot Jupiter \mbox{HD~189733b}. We analyze a single transit observed
with the high-resolution Ultraviolet and Visual Echelle Spectrograph
(UVES) at  ESO's Very Large Telescope and we apply a detailed modeling 
of the systematics. We also estimate the equivalent width of the
exoplanetary sodium line using the deformations introduced in the
excess light curves by the changing radial velocity of the planet, and
with it we estimate the abundance of sodium in the atmosphere. In
Section~\ref{sec:observation} we describe our observations and the
data reduction steps. We continue in Section~\ref{sec:data_analysis}
with a detailed description of the data analysis, the transmission
spectroscopy method, and the modeling procedure. We show our results
and the discussions in Section~\ref{sec:results}, and conclude
in Section~\ref{sec:conclusions}.

\begin{table*}[t]
\centering
\caption {Adopted values for the orbital and physical parameters of
  \mbox{HD~189733} during the fitting procedures in this work.}
\label{tbl:orbital parameters}
\begin{center}
\begin{tabular}{ l l c l }
\hline\hline
Parameter & Symbol & Value & Reference  \\
\hline
        Orbital inclination            & $i$ ($^{\circ}$)      & 85.710 $\pm$ 0.024               & \citet{Agol2010}\\
        Semi-major axis                & a ($a/R_S$)          & 8.863 $\pm$ 0.020                & \citet{Agol2010}\\
        Planet to star radius ratio    & R$_P$/R$_S$          & 0.1565                           & \citet{Sing2011}\\ 
        Mid-transit time               & T$_{0}$ (BJD$_{\text{TDB}}$) & 2454279.436714 $\pm$ 0.000015    & \citet{Agol2010}\\
        Orbital Period                 & P (days)             & 2.21857567 $\pm$ 0.00000015      &  \citet{Agol2010}\\
        Stellar effective temperature  & T$_{eff}$ (Kelvin)    & 4875 $\pm$ 43                    & \citet{Boyajian2015} \\
        Stellar surface gravity        & log g (dex)          & 4.56 $\pm$ 0.03                  &  \citet{Boyajian2015}\\
        Metallicity                    & $[\text{Fe/H}]$      &  -0.030 $\pm$ 0.08               & \citet{Torres2008}\\
        Stellar radius                 & R$_S$ (R$_{\odot}$)    & 0.756 $\pm$ 0.018                & \citet{Torres2008}\\
        Stellar rotation period        & P$_S$ (days)         & 11.953 $\pm$ 0.009               & \citet{Henry2008}\\
\hline
\end{tabular}
\end{center}
\end{table*}

%-----------------------------------------------------------------------
%                                                       METHOD
%-----------------------------------------------------------------------

\section{Observations and data reduction}
\label{sec:observation}

\subsection{Our target}

The hot Jupiter \mbox{HD~189733b} orbits a bright (V$\sim7.7$) and
active K-type star every $\sim$2.2 days. \mbox{HD~189733b} has an
atmosphere with a scale height of about \mbox{200 km}
\citep{Desert2011}. Thanks to its large scale height and bright host star,
this exoplanet has turned into one of the favorite targets for
atmospheric characterization. Especially in the case of ground-based
studies, \mbox{HD~189733b} is an excellent ``testbed'' to explore new
exoplanet atmospheric observation and characterization techniques. The
orbital and physical parameters of this system adopted in this work
are summarized in Table~\ref{tbl:orbital parameters}. To be consistent
with current data, we adopt the planet-to-star radius ratio obtained by the \citet{Sing2011} HST observations around the
sodium wavelength. However, we note that within the
precision of our data other reasonable values of \mbox{R$_P$/R$_S$}
would have produced similar results to the ones presented in this work.

\subsection{Observing log and instrumental setup}

\begin{figure}
\includegraphics[scale=0.38]{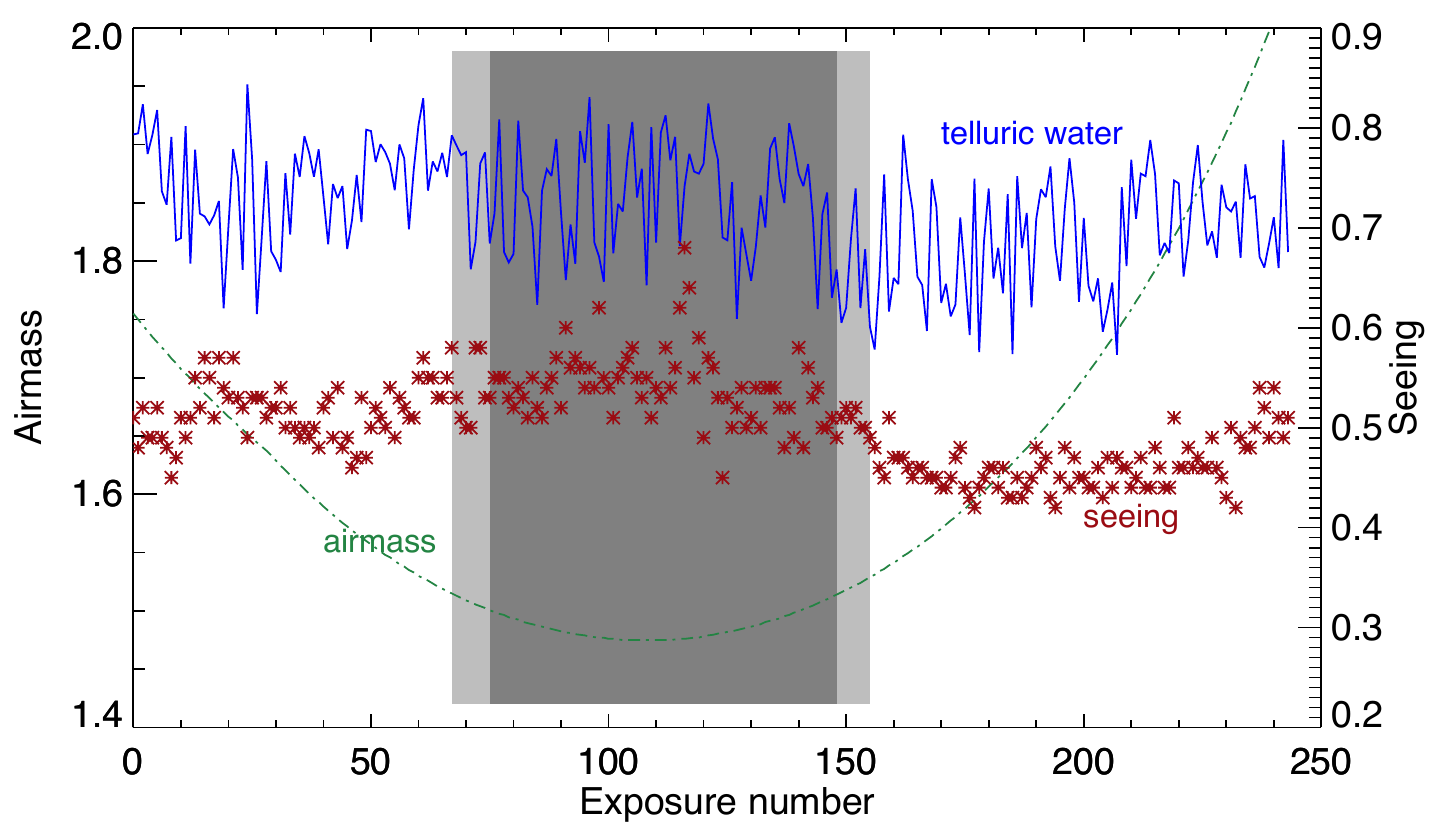}
\caption{Sky conditions during our observations. The airmass is
  indicated with a green dashed line and the red cross points indicate
  the seeing, in arcsec. The measured strength of several strong
  telluric water lines are shown with solid blue lines. These values
  are artificially shifted to fit inside the plot. Gray areas show the
  time between transit ingress and egress. Dark gray is the transit at full depth.}
\label{env-condition}
\end{figure}

During the night of July 1, 2012, a transit of \mbox{HD~189733b}
was observed using UVES \citep{Dekker2000} mounted on Kueyen, UT2
(Program ID: 089.D-0701 A). During the observations 244 spectra were
acquired: 67 exposures before ingress, 88 exposures during transit,
and 89 exposures after egress. The first 29 spectra have an exposure
time of 30 seconds, while all of the rest were  exposed for 45
seconds. The data were taken using the dichroic beam splitter with
central wavelengths around \mbox{760 nm} in the red arm and \mbox{437
  nm} in the blue arm. Our target lines are located in the red arm,
where the slit width was 0.7$^{\second}$ and the spectral resolution
is about \mbox{60\,000}. We estimate a signal-to-noise ratio around
the sodium lines of approximately 100. The sky conditions (airmass,
seeing, and telluric water) are shown in Figure~\ref{env-condition}. As
the figure shows, the depths of water lines follow a trend similar
to the seeing. However, the water lines do not clearly correlate with
airmass. We calculate the strength of the water lines by measuring
and averaging the equivalent widths of the six strongest water lines
and the values for airmass and seeing are extracted from the image
headers.

\subsection{Data reduction}

The initial data reduction and extraction of 1-D spectra were performed
using the UVES pipeline, version 5.2.0 \citep{Czesla2015}. The subsequent analysis
related to this work was conducted using Interactive Data
Language (IDL) and Python 2.7. The sequence of the reduction processes
are summarized below.

\subsubsection{Spectrum normalization}

The first step was to normalize the continuum of the spectra in the
region around the sodium lines. During the initial reduction we
removed the instrumental blaze function from each individual spectral
order. However, the spectra were still not properly normalized since
we observed variations in the continuum. To correctly normalize the spectra we first
obtained the average spectrum of all the available reduced spectra. Then
we divided this average spectrum into 60 wavelength regions within a
range of 400 $\AA$ around the sodium lines. By linearly interpolating
the maximum values within each bin we obtained the shape of the
continuum in the averaged spectrum. Finally, each individual spectrum
was divided by this interpolation.

\subsubsection{Spectrum alignment}

\begin{figure}
\centering
\includegraphics[scale=0.3]{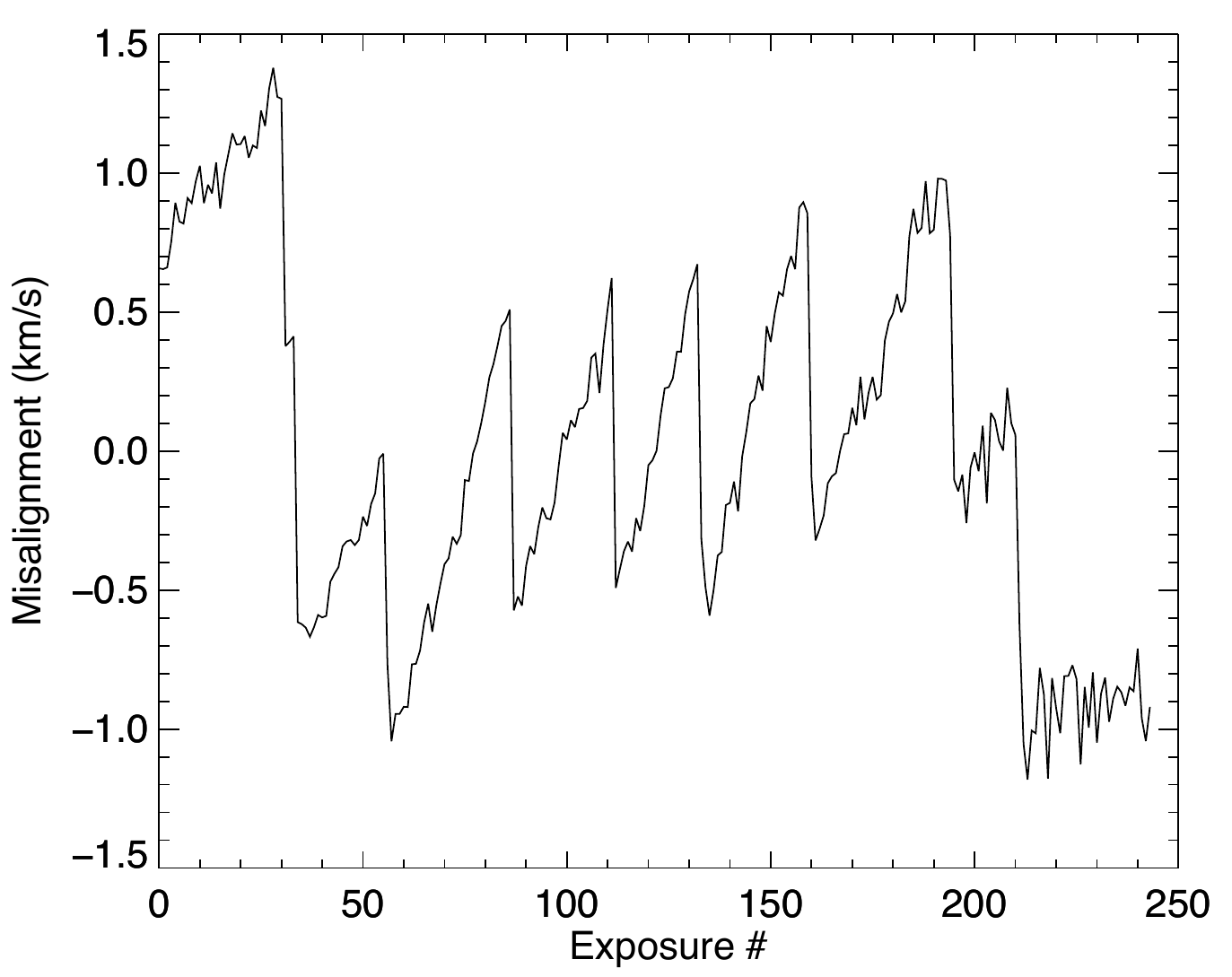}
\caption{Misalignment of each spectrum with respect to
  a reference spectrum taken in the middle of the observing run.}
\label{fig:sawtooth}
\end{figure}

Owing to the Earth's rotation and instrumental instabilities, the
wavelength solution of our spectra slightly drifts (up to about
\mbox{$\pm$ 1 km/s}). To correct for these shifts we identified 80
stellar absorption lines in a range of \mbox{400 $\AA$} around the
sodium lines. We then fitted a Gaussian profile to each one of the
line cores to obtain the position of each line center. For each line
we then calculated the time-averaged line position and thus obtained
the deviation of each line in each order of each exposure. The
strongest component is a wavelength-independent offset caused by the
entire spectrum being shifted on the detector. There may also be
wavelength-dependent deviations caused by a magnification, i.e., a
scaling of the image of the spectra on the detector, so we performed a
linear fit to the values of the misalignment of all lines in the
exposure to obtain a 2-D map of the misalignment (shift map) as a
function of wavelength. By re-interpolating all exposures to this map,
all absorption spectra were aligned, and thus placed in the same
reference frame. The amplitude of the fitted linear shift is plotted
in Figure~\ref{fig:sawtooth}, which shows that the spectra drift in
time according to a saw-tooth pattern, also identified and described
by \citet{Czesla2015}. The value of the shift in pixels rises to about
\mbox{$\pm$ 0.5 px}. To study the wavelength stability of this
dataset, \cite{Czesla2015} also obtained the radial velocity
variations of the telluric water lines in time. Since they show the
same saw-tooth behavior, we consider this to be an instrumental effect
that is likely produced by the movement of the stellar seeing disk
inside the slit. After removing the telluric water lines (see
Section~\ref{sec:telluric}), we aligned all spectra to a common
rest-frame using the shift map.

\subsubsection{Telluric correction}
\label{sec:telluric}

The Earth's telluric water and oxygen lines are removed using the
telluric absorption model described in \citet{Husser2014} and
\citet{Husser2015}. This model determines the parameters for the
stellar and Earth atmospheres simultaneously, and fits the widths and
depths of the lines. Each spectrum is then divided by its
corresponding telluric model to correct for telluric effects. It should
be noted here that visual inspection of the telluric sodium lines in
the spectra at the expected location revealed no telluric sodium
feature. In addition, the object distance is relatively close
($\sim$19 pc), which agrees with the observed lack of interstellar
sodium.

\section{Data analysis and modeling}
\label{sec:data_analysis}

\subsection{Excess light curve}
\label{sec:excess light curve}
Since the absorption cross-section of species in the atmosphere is
wavelength-dependent, the planetary radius as measured by the depth of
the transit light curve is wavelength-dependent as well. We therefore
can infer the presence of an absorbing species by obtaining the
transit light curve at that wavelength at which a given species is
expected to absorb ({integration band}), and compare it with
the transit light curve at a wavelength where the exoplanet atmosphere
is transparent (reference band). This comparison is made by
computing the ratio of the two light curves, referred to as the
``excess light curve", where the additional absorption at the
wavelength of interest as the transit progresses can be seen
\citep{Snellen2008}. In the case of sodium, we are interested in
finding the excess absorption around the optical sodium D$_{1}$ and
D$_{2}$ lines. We specifically introduce a central band

\begin{equation}
F_{\mathrm{center}} (t)= \int^{\lambda_{0}+\Delta\lambda}_{\lambda_{0} -\Delta\lambda} F(\lambda,t)d\lambda\;,
\end{equation}

\noindent where $F_{\mathrm{center}}$ describes the integrated flux
inside the target line that is expected to change in time (t) in the case
of exoplanetary atmospheric absorption, and $\lambda_{0}$ is the
central wavelength of the target absorption line. We then define left
and right reference bands

\begin{equation}
F_{\mathrm{left}} (t)= \int^{\lambda_{0}-M\Delta\lambda}_{\lambda_{0}-N\Delta\lambda} F(\lambda,t)d\lambda\;,
\end{equation}

\noindent and

\begin{equation}
F_{\mathrm{right}} (t)=\int^{\lambda_{0}+O\Delta\lambda}_{\lambda_{0}+P\Delta\lambda} F(\lambda,t)d\lambda\;,
\end{equation}

\noindent where M, N, O, and P are suitably chosen coefficients that
determine the location and width of the reference bands in the
continuum close to the sodium lines, and $\Delta\lambda$ is half
the integration band width. From $F_{\mathrm{left}}$,
$F_{\mathrm{right}}$, and $F_{\mathrm{center}}$ we construct the
relative line flux, $F_{\mathrm{line}}$, defined as

\begin{equation}
F_{\mathrm{line}}(t) = \frac{2F_{\mathrm{center}}(t)}{F_{\mathrm{left}} (t)+{F_{\mathrm{right}} (t)}}.
\label{eq:transmission_spec}
\end{equation}

\noindent The parameter $F_{\mathrm{line}}$ is the flux ratio of the integration and
reference bands, which determines the excess light curve. In this
fashion, all flux variations that affect the central and the left and
right reference bands in the same way cancel out, while variations
that affect only the central band such as absorption by the
exoplanetary atmosphere remain.

\begin{figure*}
\centering
\includegraphics[scale=0.31]{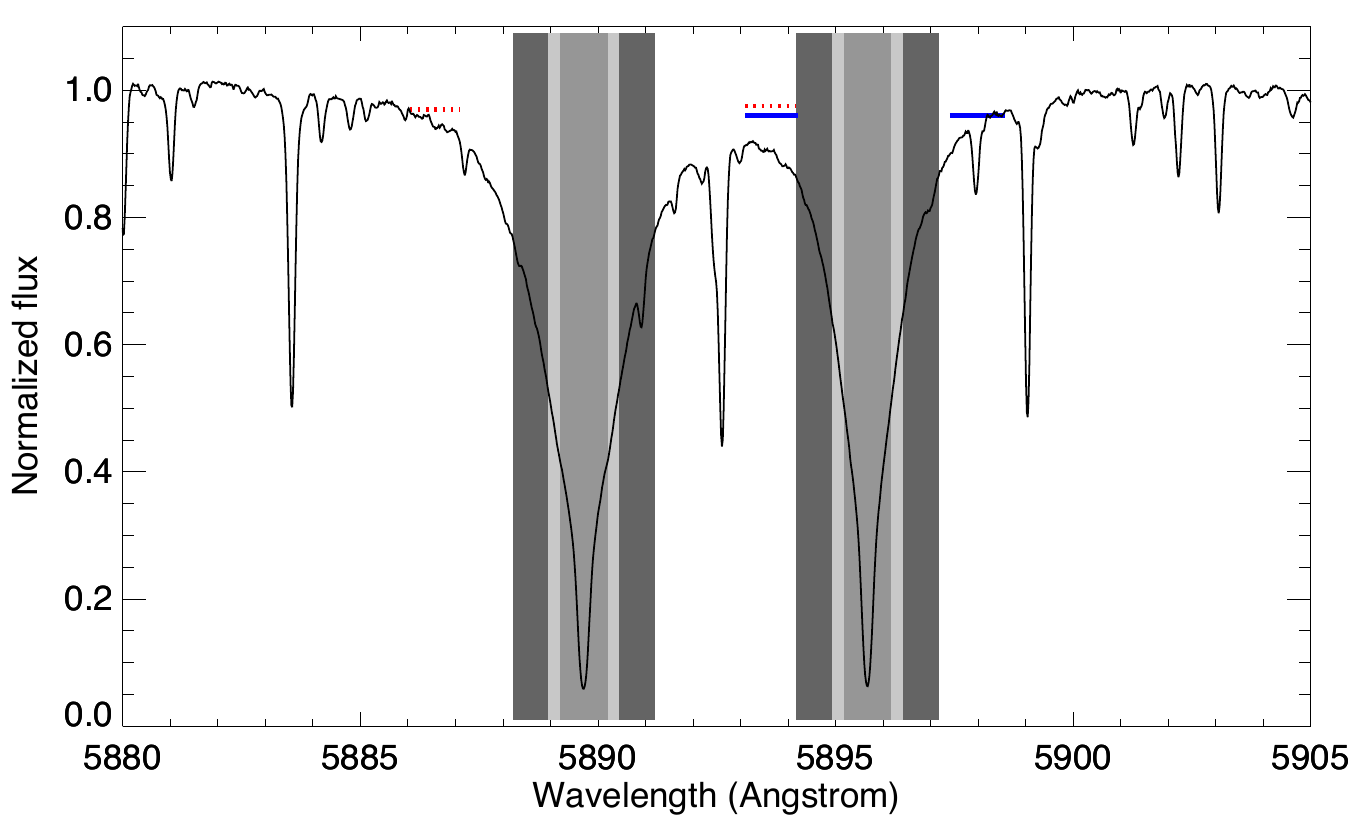}
\includegraphics[scale=0.41]{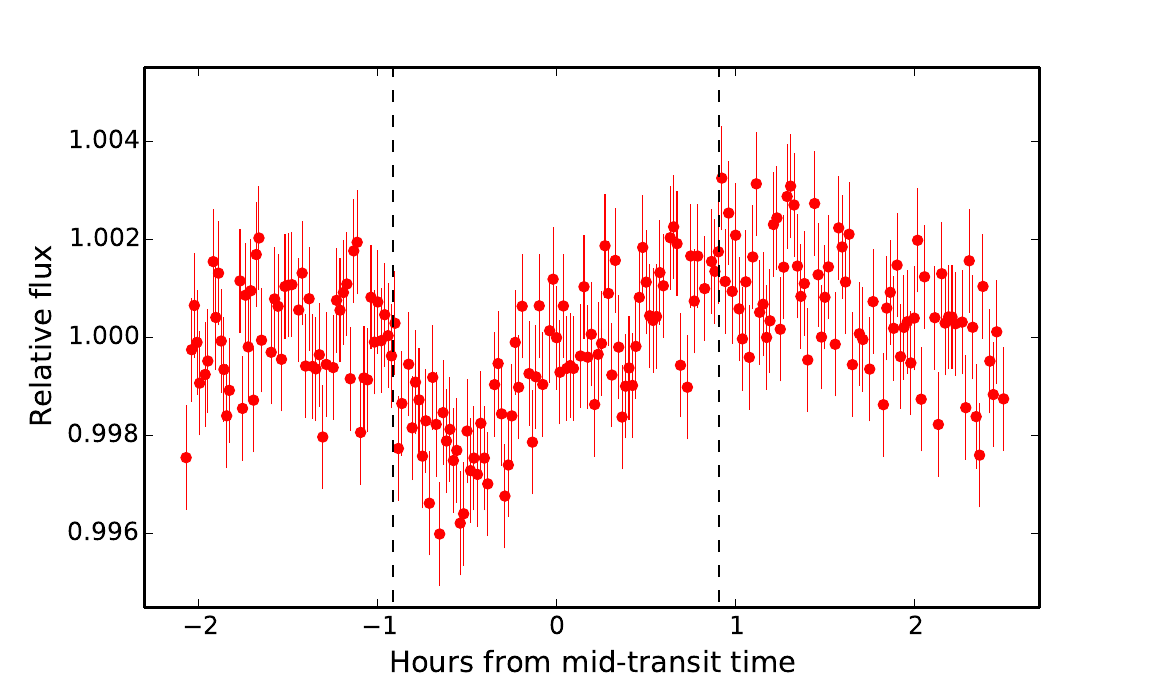}
\caption{Integration bands and derived excess light curve. \textbf{Left}:
  Gray shaded areas show the integration bands with passband of 1
  $\AA$, 1.5 $\AA,$ and 3 $\AA$, centered at the core of each sodium
  line. The red and blue intervals show the passbands of the reference
  bands for D$_{2}$ and D$_{1}$, respectively. \textbf{Right}: The raw
  excess light curve shown with the red circles for the 1.5~$\AA$
  integration band inside the sodium D$_{2}$ line. Dashed vertical lines
  indicate times of ingress and egress.}
\label{shaded}
\end{figure*}

In our analysis we first locate the centroids of each sodium D-line
in every reduced spectrum by fitting a Voigt profile. Then we chose
flux integration passbands of \mbox{1 $\AA$}, \mbox{1.5 $\AA$}, and
\mbox{3 $\AA$} centered on each line to characterize the extra
\mbox{Na I} absorption from the planetary atmosphere. The smallest
passband width is set by the signal-to-noise ratio in the deep cores
of the \mbox{Na I} lines; the aperture needs to be wide enough to
include enough photons to be sensitive to small variations in the
depth of the \mbox{Na I} lines. The upper limit in the width of the
passband is given by the value at which the planetary \mbox{Na I}
signal blends with the noise. Specifically, our \mbox{3 $\AA$}
integration aperture upper limit is determined empirically by
comparing the standard deviations in the light curves for each
passband to that  of the reference continuum. The standard deviation of
the reference continuum light curve, where one does not expect to have
any planetary atmospheric signal, is 0.83 parts per thousand (ppt),
which we attribute to random noise. In the case of the \mbox{Na I}
line passbands, the standard deviations of the light curves are 1.9
ppt, 1.4 ppt, and 0.85 ppt for the \mbox{1 $\AA$}, \mbox{1.5 $\AA$},
and \mbox{3 $\AA$} passbands, respectively. These values clearly
indicate the presence of an additional signal in the \mbox{1 $\AA$}
and \mbox{1.5 $\AA$} passbands, while the noise level of the \mbox{3
  $\AA$} passband is similar to that of the reference continuum,
placing the signal and the noise at the same level. The three
integration bands are illustrated in Figure~\ref{shaded}, left,
together with the sections of the spectrum we used as reference
continuum. For a comparative analysis of our results with those of  other
authors, our integration passbands are close to the integration
passbands of \citet{Snellen2008}, \citet{Albrecht2008Thes}, and
\citet{Wyttenbach2015}. Our reference bands are fixed to the same
values for all the integration bands, and are placed closer to the
upper wings of the sodium lines to minimize the difference between the
limb-darkening (LD) coefficients in the integration and reference bands
(see Section~\ref{sec:LD}). We also  place them in a location
free of strong absorption features. Taking all of this into account,
our reference passbands have a width of 1$\AA$ and are placed on each
side of each sodium line, as indicated in Figure~\ref{shaded}. The
reference bands between the lines are taken to be the same for both
sodium D-lines, otherwise for the D$_{2}$ line a fraction of the
reference band would fall inside the deep iron line in between the
sodium lines.

Finally, we use Equation~\ref{eq:transmission_spec} to obtain the
relative strength of the sodium absorption features in each
exposure. The excess light curves we obtain by using this
approach on the sodium D$_{2}$ line for the integration bands of \mbox{1.5
  $\AA$} are shown in Figure~\ref{shaded}, right, and we call it
the ``raw excess light curve". As the figure shows, the raw
excess light curve does not look transit-like and is probably highly
affected by external physical, environmental, or instrumental effects. In the next sections we investigate the
source of these effects and use them to produce a detailed model
fit to the data.

\subsection{Model components and parameters}
\label{sec:model_parameters}

Before drawing any conclusions on the \mbox{HD~189733b} atmosphere,
the external effects present in the raw excess light curves have to be
taken into consideration. In total, we identify three main effects:
 the occurrence of a stellar flare (\citealt{Czesla2015} and Klocov\'a et al., submitted to A\&A) starting close
to mid-transit time (A);   a wavelength-dependent limb-darkening (B); and   the observed changes in the line profile
produced when the planetary sodium line moves inside the stellar
sodium line, induced by the planetary orbital motion
(C). In the next
sections we  explain each effect separately and  specify their
fitting parameters. The magnitudes of the individual errors for the
excess light curves are determined in two ways,  by computing the
standard deviation of data points between the 30th and 100th
exposures, and by computing the standard deviation of the residual
light curves  produced by subtracting a least-squares fit of
the four model components arranged as in Equation~\ref{eq:model} to
the data. For a more conservative approach, the values of the errors
are assigned by choosing the largest value between the two approaches.

\subsubsection{Stellar flare (component A)}
\label{sec:flare}

\mbox{HD~189733} has been characterized as a highly active K-type star
\citep[e.g.,][]{Poppenhaeger2013}. As a result, our observations are
prone to being contaminated by stellar flares and/or spots. Indeed, we
detect a rising pattern of flux that starts right after the
mid-transit point, identified as a stellar eruption. Usually, a clear
evidence of a stellar flare can be derived from the measurements of
the equivalent widths of emission lines in the cores of \mbox{Ca II} H
and K and H$\alpha$ lines originating in the chromospheric layers
(e.g., Klocov\'a et al., submitted to A\&A). By analyzing these spectral lines, we
confirmed the presence of a flare, which -- as predicted -- effectively took
place close to mid-transit time \citep[see top panel of Figure 11
  in][]{Czesla2015}. Although this feature is identified using the
\mbox{Ca II} H and K lines, stellar flares might affect other spectral
lines. For instance, neutral sodium in the lower stellar atmosphere
can be contaminated by this kind of eruption
\citep{Cessateur2010}. Therefore, in our specific case of study it is
important to consider this time-dependent change in the core of the
line as part of our model budget. To mitigate the flare, we made use of
the variations in the equivalent width of the \mbox{Ca II} K line (Klocov\'a et al., submitted to A\&A), and
we call this model component the flare profile,
$\text{flr}_{\text{Ca}}(t)$. To quantify to what extent the
flare correlates with our raw sodium excess light curve, we used a
Pearson's correlation analysis. The six computed values of Pearson's
coefficient, computed from the two sodium lines times three integration
bands, range between 0.25 and 0.45, and their corresponding p-values
are in five of the six cases smaller than 10$^{-3}$. For the \mbox{3
  \AA} integration band and the D$_{2}$ line, the p-value is 0.02. In these
cases, the computed p-values indicate the probability of an
uncorrelated system producing data sets that have a Pearson
correlation at least as extreme as that computed from these data
sets. In other words, they indicate that the null hypothesis, which
states that the raw excess light curve and the flare are uncorrelated,
can be rejected with high significance. Thus, there is enough
statistical evidence that the flare has to be taken into
consideration. To mitigate the stellar activity present in our excess
light curves, we fitted  a scaled version
of the \mbox{Ca II} K flare profile to the raw excess light curves with the  expression

\begin{equation}
\text{flr}_{\text{Na}}(t) = \text{FL}_{\mathrm{scale}}\cdot\text{flr}_{\text{Ca}}(t)\;,
\label{eq:flare}
\end{equation} 

\noindent where FL$_{\mathrm{scale}}$ is the fitting parameter that
scales the flare, and flr$_{\text{Na}}$ is the model component that
accounts for the stellar flare. Our best-fit model for the Na D$_{2}$ line
at \mbox{1.5 \AA} integration width, showing its isolated flare
component, can be seen in \mbox{Figure~\ref{ABCD}-A} as a solid black
line, plotted along with the raw excess light curve in red points.

\subsubsection{Wavelength dependent limb-darkening effect (component B)}
\label{sec:LD}

\begin{figure*}
\centering
\includegraphics[width=1.0\textwidth]{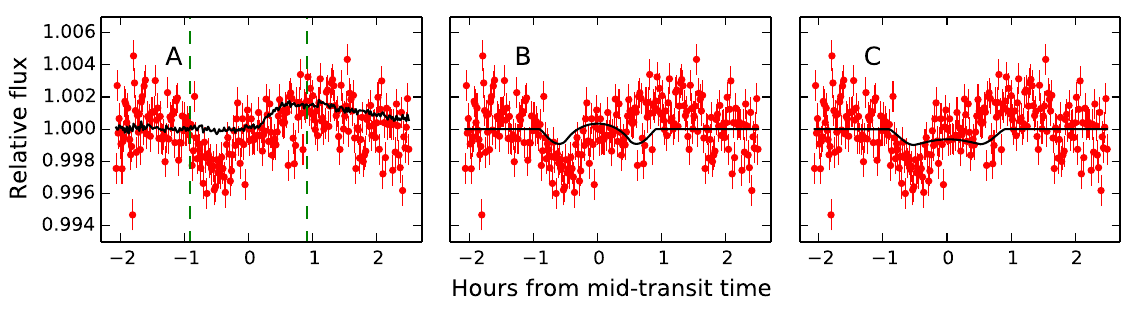}
\caption{Individual model components plotted over the excess
  light curve for Na D$_{2}$ at 1.5$\AA$. \textbf{Panel A:} Best-fit flare model component. In this panel the green dashed lines indicate the beginning and the end of the transit; \textbf{panel B:} Best-fit differential
  limb-darkening model component; \textbf{panel C:} Best-fit
  changing RV model component, which includes the exoplanetary
  atmospheric excess absorption in addition to a bump at the
  center. The final best-fit model, which is the combination of all the
  model components, is shown in Figure \ref{fig:raw_model_res}.}
\label{ABCD}
\end{figure*}

\begin{table*}[ht!]
\caption {Limb-darkening coefficients (u$_{1}$,u$_{2}$) for the
  sodium D$_{2}$ and D$_{1}$ integration bands (Core) and fixed
  reference (Ref.) bands with the specified wavelength ranges
  ($\lambda$-range). Errors of the limb-darkening coefficients are not
  shown to avoid visual contamination, but are in all cases on the
  order of 10$^{-3}$-10$^{-4}$.}
\label{tbl:LDsD2-D1}
\begin{center}
\begin{tabular}{ c | c c c c c }
\hline\hline
\textbf{D$_{2}$ Core/Ref.}        & 1 $\AA$ left fixed ref. & 1 $\AA$ core        & 1.5 $\AA$ core     & 3 $\AA$ core         & 1 $\AA$ right fixed ref.\\
\textbf{$\lambda$-range ($\AA$)} & [5886.04 , 5887.10]     & [5889.42 , 5890.48] & [5889.19 , 5890.71] & [5888.49 , 5891.41] & [5893.38 , 5894.44]\\
\textbf{u$_{1}$,u$_{2}$}          & 0.6856, 0.0760           & 0.6051 , 0.0149     & 0.5961 , 0.0318    & 0.5987 , 0.0556     & 0.7078 , 0.0187 \\
\hline
\textbf{D$_{1}$ Core/Ref.}        & 1 $\AA$ left fixed ref.  & 1 $\AA$ core       & 1.5 $\AA$ core     & 3 $\AA$ core        & 1 $\AA$ right fixed ref.\\
\textbf{$\lambda$-range ($\AA$)} & [5893.38 , 5894.44]      & [5895.40 , 5896.45] & [5895.17,5896.68] & [5894.46,5897.39]   & [5897.51 , 5898.57]\\
\textbf{u$_{1}$,u$_{2}$}           & 0.7078 , 0.0178          & 0.6813 , 0.0081    & 0.6244 , 0.0477    & 0.6300, 0.0606      & 0.6649 , 0.0947 \\
 \hline
\end{tabular}
\end{center}
\end{table*}

Stellar limb darkening is wavelength-dependent. Thus, the
limb-darkening coefficients inside the broad and deep lines are
different from the neighboring continuum. Since the construction of
the excess light curve relates the core of the Na D-lines to their
near continuum, the difference in limb-darkening values will have an
impact on the shape of the excess light curve \citep[see,
  e.g.,][]{Charbonneau2002,Sing2008,Czesla2015}. We can formulate this
effect using the  relation
  
\begin{equation}
LD = \frac{2\times \mathrm{LC}_{\mathrm{core}}}{\mathrm{LC}_{\mathrm{left}} + \mathrm{LC}_{\mathrm{right}}}\;,
\label{eq:LD}
\end{equation} 

\noindent where LD is the wavelength dependent limb-darkening model,
LC$_{\mathrm{core}}$; LC$_{\mathrm{left}}$ and LC$_{\mathrm{right}}$
are the transit light-curve models from \cite{Agol2002} using the
orbital parameters tabulated in Table~\ref{tbl:orbital parameters},
 which have their own limb-darkening coefficients at the core of the
sodium line, and at the left and right reference passbands,
respectively \citep[see the limb-darkening model in
  Figure~\ref{ABCD}-B and Figure~2 of][]{Czesla2015}. To take this
effect into consideration we first computed the limb-darkening
coefficients in all relevant (integration and reference) passbands
using the PHOENIX angle-resolved synthetic spectra
\citep{Hauschildt1999,Husser2013} implementing the method detailed in
\cite{vonEssen2013}. PHOENIX requires effective temperature, surface
gravity and metallicity of the star as input, for which we adopted the
values \mbox{T$_{eff}$ = 4900 K}, \mbox{[Fe/H] = 0}, and \mbox{log$g$
  = 4.5}, closely matching the stellar parameters of \mbox{HD~189733}
listed in this work in Table~\ref{tbl:orbital parameters}. The derived
angle-dependent intensities were then fitted with a quadratic
limb-darkening law,

\begin{equation}
\frac{I(\mu)}{I(0)} = 1- u_{1}(1-\mu) - u_{2}(1-\mu)^{2}\;,
\label{eq:limb}
\end{equation}

\noindent where I($\mu$)/I(0) corresponds to the normalized stellar
intensity,  $u_1$ and $u_2$ correspond respectively to the linear and quadratic limb
darkening coefficients, and $\mu$ is $\cos\theta$, where $\theta$ is the angle between the
normal to the stellar surface and the line of sight to the
observer. In this fashion, \mbox{$\mu$ = 0} refers to the limb, while
\mbox{$\mu$ = 1} corresponds to the center of stellar disk. To carry
out the fit between the integrated stellar intensities and
Equation~\ref{eq:limb}, we consider the $\mu$ values between
0.1 and 1 to avoid the steep intensity gradient appearing close to the
stellar limb that, if accounted for, would have to be fitted with an
added exponential growth to the quadratic function. Since the
orientation of the \mbox{HD~189733b} orbit does not produce a nearly
grazing transit, the time that the planet spends in the 0-0.1
$\mu$-region is very short and our simplification is justified. The
derived values for the sodium D$_{2}$ and D$_{1}$ lines in their
corresponding integration and reference bands are listed in
Table~\ref{tbl:LDsD2-D1}. 
Here it should be noted that since the integration
band widths are, in this work, between 1 and 3 \AA, we re-computed the PHOENIX spectra, increasing the
spectral sampling to \mbox{$\sim$0.1 \AA}.

As previously mentioned, using the computed limb-darkening values we
produce three light-curve models \citep{Agol2002} and calculate the
wavelength dependent limb-darkening model introduced in
Equation~\ref{eq:LD}. The sensitivity of our data is not high enough
to consider the limb darkening coefficients as fitting
parameters. Therefore, this model is taken into account but with all
their values fixed. The limb-darkening model is shown in
Figure~\ref{ABCD}-B as a  black line, overplotted onto the raw excess
light curve.

\subsubsection{Planetary radial velocity (component C)}
\label{sec:bump}

\begin{figure}[ht!]
\centering
\includegraphics[width=0.5\textwidth]{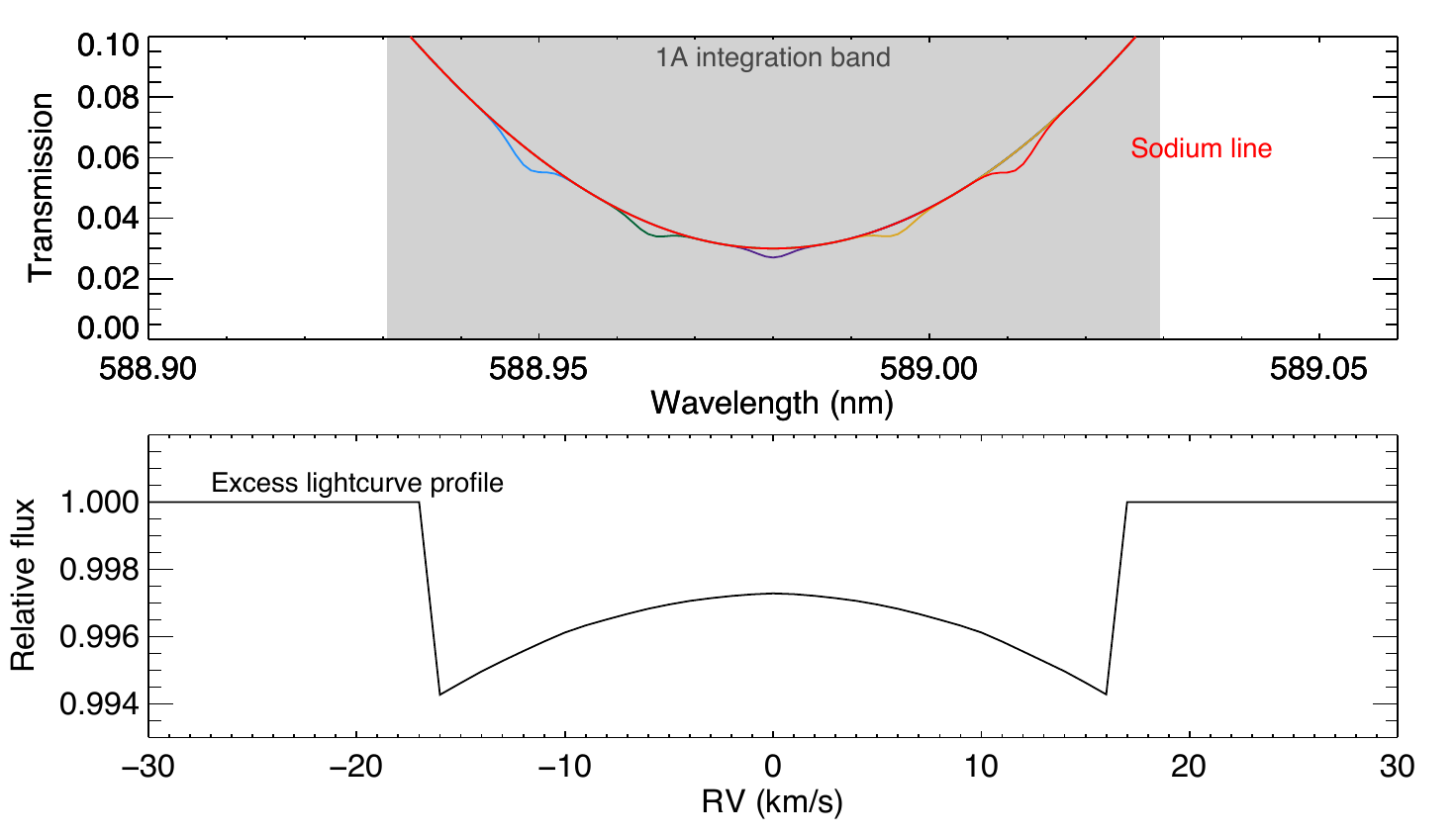}
\caption{\textbf{Top}: Schematic shift of the planetary sodium line
  inside the stellar sodium line during the transit. The size of the
  planetary line has been exaggerated. The gray area corresponds to
  \mbox{1 \AA} integration band. \textbf{Bottom}: Effect of the
  planetary line shift on the excess light curve with the integration
  passband of \mbox{1 \AA}, causing a bump that reaches its maximum
  level around the mid-transit point. We note that the x-axis is also
  proportional to the transit time, with zero set on the mid-transit
  point.}
\label{fig:bump} 
\end{figure}

Because the planet is moving during its transit on a curved orbit, its
radial velocity changes. Hot Jupiters often have orbital speeds on the
order of \mbox{100 km/s}, which can result in an in-and-out transit
radial velocity variation larger than \mbox{$\pm$10 km/s}. This change
in radial velocity causes the entire transmission of the exoplanet to
be Doppler shifted, and has already been used to detect molecules in
the atmospheres of hot Jupiters \citep[e.g.,][using very high-resolution spectra,
  R$\sim$100\ 000]{Snellen2010,Brogi2012,Rodler2012}. In our case as
illustrated in Figure \ref{fig:bump}-top, the sodium line originating
in the planetary atmosphere shifts between the lower wing and the core
of the stellar sodium line. During this shift, the planet atmosphere
always absorbs a certain fraction of starlight. In other words, the relative absorption of stellar flux by the planet atmosphere is always the same. This causes
the excess light curve to increase at transit center; when the stellar
flux is at minimum level, the planetary sodium line is close to
the center of the core of the stellar line and the radial velocity of the
planet is close to 0. Near ingress and egress, the planetary sodium
line is not located in the direct vicinity of the stellar line
core. The planetary sodium line at these times is embedded in a part
of the stellar line wing where the stellar flux is higher than in 
the core. Here it absorbs the same fraction of starlight, but this
absorption is larger in absolute terms. This causes a ``bump" in
the bottom of the transit, which is illustrated in Figure
\ref{fig:bump}-bottom. This effect was first mentioned by
\citet{Albrecht2008Thes} and can also be tentatively seen in the results of the sodium core analysis by
\citet{Snellen2008}, \citet{Zhou2012}, \citet{Wyttenbach2015}, and \citet{Cauley2016}. In
the particular case of \mbox{HD~189733b}, during transit the radial
velocity of the exoplanet changes between \mbox{-16 km/s} and
\mbox{+16 km/s}. This results in a Doppler shift of approximately
\mbox{$\pm$0.31 $\AA$} around the core of the stellar sodium
line. We note that with this amount of shift, the exoplanetary sodium
absorption line is always embedded in the stellar sodium line of the K-type. To
model the radial velocity effect, we time-averaged all out-of-transit
spectra to obtain the shape of the stellar sodium lines at high
signal-to-noise, and used this as the stellar sodium line template. We
represented the planetary line as a Gaussian profile with depth,
A$_{\mathrm{Na}}$, and width (Gaussian sigma), $\sigma_{\mathrm{Na}}$, both considered
in this work as fitting parameters. This Gaussian profile is injected
into the time-averaged sodium lines calculating the expected radial
velocity of the planet ($V_{R}$) during each exposure (see
Equation~\ref{eq:RV}). In other words, we create 244 spectra by
multiplying the moving planetary sodium line Gaussian profiles by the
time-averaged normalized stellar spectrum. We call this new series of
spectra the ``convolved spectra". The radial velocity of the planet at each exposure is calculated as
 
\begin{equation}\label{eq:RV}
\Delta V_{R} = \frac{2~\pi~a}{P} \times \sin [2~ \pi~ \left( \frac{t-t_{c}}{P}\right)]\;,
\end{equation}

\noindent but is converted into wavelength
shift to position the center of the Gaussian profile over the stellar
sodium template. Here, $P$ is the orbital period, $a$ is the semi-major axis, $t$
is the time of observation for each frame, and t$_{c}$ is the
mid-transit time. The convolved spectra were then integrated
following equation \ref{eq:transmission_spec}. The result is a
numerical model that we call the ``initial RV model". In the initial
RV model, even the exoplanetary Gaussian profiles taking place outside
transit are also taken into account. However, in transmission
spectroscopy the planetary atmosphere is invisible before ingress and
after egress. To extract the time span in which the transit takes place
and also to correct for the shape of the RV model during ingress and
egress, the initial RV model is multiplied by a transit model that
has an offset of zero and  a normalized depth of one. The result
gives us the final RV model. Figure~\ref{fig:bump} is a mock model that
illustrates the changing radial velocity effect on the excess
light curve for an integration passband of \mbox{1 $\AA$}. In the
figure, the planetary and stellar sodium lines are both represented by
Gaussian profiles. Finally, our RV model, as a component of the
best-fit model, is plotted over the data in Figure~\ref{ABCD}-C. We note
that this model is not completely symmetric since the two stellar sodium
lines are  not fully symmetric.

\subsection{Final model and best-fit parameters}

The overall model we use to fit the data must be a combination of the
model components just introduced (A, B, C) plus a
normalization constant (offset or D) for the flare
component. None of the model components can reproduce the data when
treated individually, thus a combination is required. We
choose a model consisting of a multiplication of all of the
components plus the offset. The final model is shown in equation
\ref{eq:model}. Since we are dealing with very small effects, the
multiplication or summation of the components would both give the same results (see also  \citealt{Czesla2015}):

\begin{equation}\
\begin{array}{ll}  % lcl second line in center
\label{eq:model}
M(t_{i}) & = \text{flr}_{\text{Na}}(t) \times LD(t) \times RV(t) + offset\\
& \equiv A \times B \times C  + D\\
\end{array}
\end{equation}

The final model (Equation~\ref{eq:model}) has four fitting
parameters in total: the flare scaling parameter (FL$_{\mathrm{scale}}$), the amplitude
(A$_{\mathrm{Na}}$) and width ($\sigma_{\mathrm{Na}}$) of the
planetary sodium Gaussian profile, and the normalization constant
(offset). Throughout  this work, to explore the best-fit parameters
and their associated uncertainties we perform a Markov chain Monte
Carlo (MCMC) analysis, using the affine invariant ensemble sampler
\emph{emcee} \citep{Foreman2013}. We employ 100 walkers, with 360
chains each, where the initial positions are synthesized from a
Gaussian distribution about our best estimates. All the free parameters have uniform prior imposed. We allow a burn-in
phase of $\sim 50\%$ of the total chain length, beyond which the MCMC
is converged. The posterior probability distribution is then
calculated from the latter 50\% of the chain.

%---------------------------------------------------------------------------------------------
%                               RESULTS
%---------------------------------------------------------------------------------------------

\section{Results and discussion}
\label{sec:results}

\subsection{Best-fitting model and errors}
\label{sec:best_model}

\begin{figure*}
\centering
\includegraphics[width=.8\textwidth]{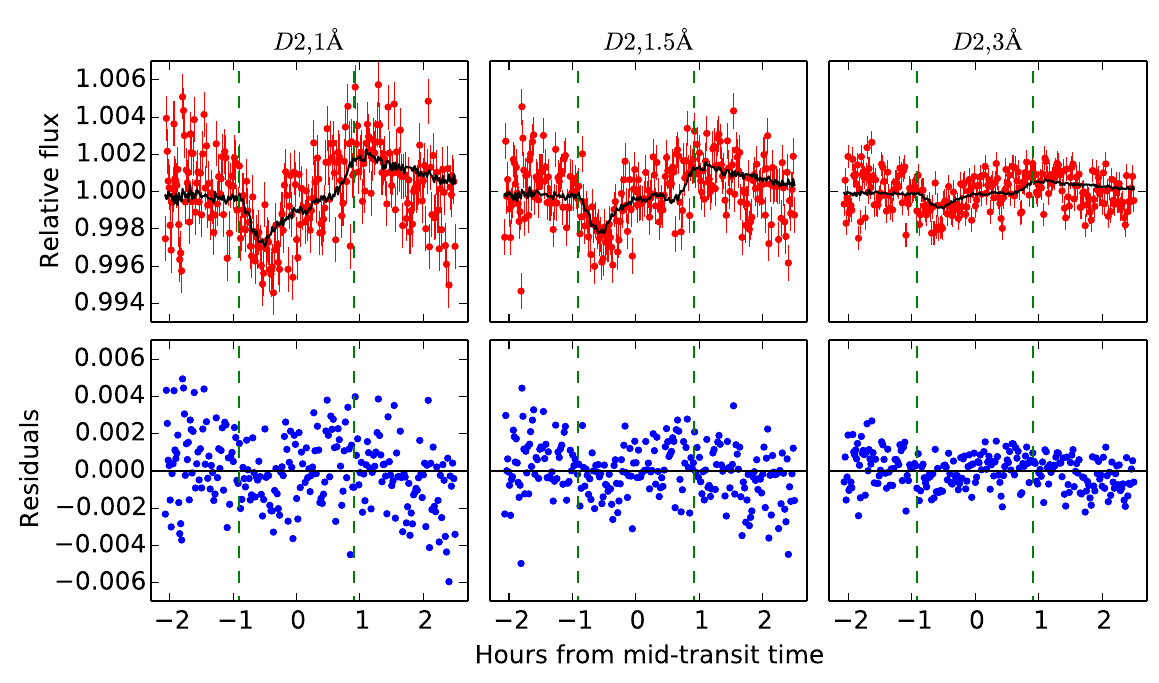}
\caption{\textbf{First row:} Raw excess light curves on sodium D$_2$
  for integration passbands of \mbox{1 \AA}, \mbox{1.5 \AA}, and
  \mbox{3 \AA} inside the core of each sodium line. The best-fit model
  is plotted on top in black. \textbf{Second row:} Residuals for each
  passband  D$_1$.}
\label{fig:raw_model_res}
\end{figure*}

\begin{table*}[ht!]
\centering
\begin{tabular}{c |c  c | c c}
\hline\hline
Passband & FL$_{\mathrm{scale}}$ & offset  &  $\sigma_{\mathrm{Na}}$  &  A$_{\mathrm{Na}}$ \\
\hline
D$_{1}$, 1$\AA$ & 0.020 $\pm$ 0.002  & 0.00021 $\pm$ 0.00008   & 0.09 $\pm$ 0.02  & 0.024 $\pm$ 0.005  \\
D$_{1}$, 1.5$\AA$ & 0.014 $\pm$ 0.002  & -0.00005 $\pm$ 0.00005  & 0.07 $\pm$ 0.02  & 0.018 $\pm$  0.006  \\
D$_{1}$, 3$\AA$ & 0.007 $\pm$ 0.001  & -0.00002 $\pm$ 0.00002  & 0.03 $\pm$ 0.02  & 0.009 $\pm$ 0.007 \\
\hline
D$_{2}$, 1$\AA$ & 0.046 $\pm$ 0.003  & - 0.0003 $\pm$ 0.00007  & 0.10 $\pm$ 0.02 & 0.036 $\pm$ 0.011 \\
D$_{2}$, 1.5$\AA$ & 0.032 $\pm$ 0.002  & - 0.0002 $\pm$ 0.00006 & 0.13 $\pm$ 0.04 & 0.013  $\pm$ 0.005 \\
D$_{2}$, 3$\AA$ & 0.014 $\pm$ 0.002 & - 0.0001 $\pm$ 0.00006  &  0.04 $\pm$ 0.02 & 0.022 $\pm$ 0.015 \\
\hline
\end{tabular}
\caption {\label{tbl:best-fit-values} Best-fit values for the model parameters obtained from the
  Na D$_{1}$ and Na D$_{2}$ excess light curves in
  the three integration bands. As a reminder, here \textbf{FL\boldmath$_{\mathrm{scale}}$} is the flare scaling parameter, \textbf{offset} is the normalization constant for flare models, \boldmath$\sigma_{\mathrm{Na}}$ is width of the exoplanetary Gaussian profile, and \textbf{A$_{\mathrm{Na}}$} is the amplitude of the exoplanetary Gaussian profile.}
    %To allow a better visualization, parenthesis at the end of each value indicate the decimals of precision of the derived errors
\end{table*}

Our best-fitting values of the MCMC procedure are summarized in
Table~\ref{tbl:best-fit-values}. To illustrate the quality of our fit,
Figure~\ref{banana} shows the posterior distributions and the
correlation plots for all the fitting parameters, computed here as an
example for the \mbox{1.5 \AA} excess light curve around the sodium D$_{2}$
line. The five remaining excess light curves produce similar plots.
%Except for the strong correlation between AM$_{\mathrm{scale}}$
%and OFF, values of the Pearson's correlation coefficient between 0.05
%and 0.3 for pairs of parameters seem to indicate weak correlation
%between them.
As the posterior distributions in Figure \ref{banana} show, the
$\sigma_{\mathrm{Na}}$ and A$_{\mathrm{Na}}$ parameters are degenerate
and the rest of parameters are not correlated at all or only very weakly. However, in this work we use them to calculate the
equivalent width of the sodium line (equivalently an area, nothing
more than a direct multiplication of both parameters). Therefore, the
degeneracy between parameters does not affect our results.

The best-fit model for sodium D$_{2}$ passbands and the
corresponding residuals are shown in
Figure~\ref{fig:raw_model_res}. As the figure shows, the transit
depth, the flare profile amplitude, and the bump strength decrease
when the integration passbands increase.
In smaller passbands, the planetary signal is more pronounced because there is a higher contrast between the exoplanetary absorption and stellar
absorption. Moreover, in smaller integration passbands fewer points are added together and thus -- compared to wider passbands where the random noise cancels out better -- the light curve is more scattered.

To ensure that equation \ref{eq:model} is the best representation of
the data, we performed some statistical tests.  We made use of the
Bayesian Information Criterion \citep{Schwarz1978}, \mbox{BIC = $\chi^2$ + k ln N}, where
k is the number of model parameters and N is the number of data
points. The BIC assess model fits penalized for the number of
estimated parameters. We also compute chi-square ($\chi^{2}$) and
reduced chi-square ($\chi^{2}/\nu$), being $\nu$ the number of degrees
of freedom equal to the total number of data points minus the number
of fitting parameters for each model combination. Producing this
statistical analysis for all the possible combinations of model
components would not be very efficient since some of the possible
combinations are trivially not the best-fit answer. Therefore, we
exclude some of them and pre-selected five different model combinations
(modules) using visual inspection as a criterion, and calculated the
previously mentioned parameters for each one. Our results for
the passband of \mbox{1.5 $\AA$} is shown in
Table~\ref{tbl:stat_test}. Tests on other passbands also
showed similar results.  Based on the values of the table, the models
without the flare (component A) and exoplanetary atmosphere absorbing
signal (component C) are very poor. Without
considering component B, the $\chi^2$, $\chi^{2}/\nu$ and BIC values
did not change much and even slightly decreased. It is important to keep in mind 
that the differential limb-darkening is a physical effect that
inevitably exists. Without this effect the exoplanetary signal must be
overestimated. Therefore, although the goodness of the fit did
not change much, component B was also included in defining the best
model combination. Thus, briefly, from the minimization of the three
statistical tools, we conclude that a good representation of the data
is given by a combination of the model components A, B, and C plus
the normalization constant D. The final models are introduced in
equation~\ref{eq:model} and the best-fit model for all of the
passbands are shown in Figure \ref{fig:raw_model_res}.
As can be seen in this figure and  in Figure \ref{ABCD}-A, near the end of the observation the flare model does not perfectly fit the data. One probable reason is the shape of the flare at Ca II K lines is slightly different from the shape in sodium lines.

\begin{table*}[ht!]
\center \caption{Statistical tests for some model combinations for the
  \mbox{1.5 $\AA$} passband.}
\label{tbl:stat_test}
\begin{tabular}{ c c | c c c c c }
\hline\hline
Test  & Module & *k &  *N & $\chi^{2}$ & $\chi^{2}/\nu$ & BIC \\
\hline
1 &     A$\times$B$\times$C + D      & 4     &  244 & 312 & 1.29 & 333 \\
2 &     A$\times$B + D       & 2     &  244  & 330 & 1.36 & 341 \\
3 &     A$\times$C + D       & 4     & 244 & 308 &  1.29  &  331 \\
4 &     B$\times$C + D       & 4     & 244  & 365  & 1.52  &  382 \\
5 &     A + E       & 2     & 244  &  356 & 1.47 &  367 \\
\hline
\end{tabular}
%\\[2]
\begin{tablenotes}
\item \center *\tiny k is the number of model parameter and N is the
  number of data points. $\nu$ denotes the degrees of freedom, being
  $\nu$ = N - k.
\end{tablenotes}
\end{table*}

Furthermore, we  tested whether         assuming a constant model for the limb darkening effect has an impact on our results. A change in the LD model would be the product of miscalculated limb darkening values. Since there is no empirical way to test the accuracy of the LD values, we carried out the following test: Any change in the limb darkening values would mostly change the amplitude of the two features centered around the mid-transit point, either by increasing or decreasing it. Therefore, we ran our whole fitting algorithm two more times, arbitrarily increasing this amplitude by 50\% and decreasing it by the same amount with respect to its original shape. After computing the excess depths in the exact same fashion as explained in this work but with the LD model component changed, we observed no significant difference between different results when 1$\sigma$ errors are being considered. Thus, the precision of our data does not allow for a limb darkening fitting.

\subsection{Exoplanetary sodium line}

Similar to \cite{Albrecht2008Thes} and \cite{Zhou2012}, our data
clearly show a reduction of the exo-atmospheric absorption in the
middle of the transit which is caused by the radial velocity shifts of
the planetary sodium signal inside the broad stellar sodium line. We
use this effect to model the sodium D-lines in the atmosphere of
\mbox{HD~189733b}. According to our model parameters, we estimated the
planetary sodium line depth, A$_{\mathrm{Na}}$, and a measurement of the line
width, $\sigma_{\mathrm{Na}}$. Our best-fit values, along with 1$\sigma$
errors, are summarized in the   last two  columns of
Table~\ref{tbl:best-fit-values} for both planetary sodium D$_{1}$ and
D$_{2}$ lines in the three integration bands. We find the values of
the line depth and line width for the two  narrower passbands to
be consistent with each other within the given errors. Obviously this
should be the case, since the shape of the planetary signal moving
inside the stellar sodium line should be independent of the choice of
passband. However, amplitudes and depths at 3 $\AA$ are slightly below
the error bars of the other two passbands. 
As we mentioned before (in section \ref{sec:excess light curve}), with this dataset at 3 $\AA$ the quality of the data is insufficient most probably due to the flare and short cadence of the data. In here, that must be the reason for very large error bars and underestimated planetary signal. 
Therefore, when calculating  the average of the $\sigma_{\mathrm{Na}}$
and A$_{\mathrm{Na}}$, we exclude the results of the 3 $\AA$ data. The average line
depth and width are $\bar\sigma_{\mathrm{Na}} \simeq (0.098 \pm 0.026)$ $\AA$ and $\bar
A \simeq 0.023 \pm 0.007$. We also estimated the equivalent width to be
the product of the averaged line depth and the averaged line
width, which is $\sim (0.0023 \pm 0.001)$ $\AA$.
%Calculating the equivalent widths for sodium D$_{1}$ and
%D$_{2}$ individually, shows that D$_{2}$ is larger than D$_{1}$ by a
%factor of 2.

In principle, the width of the line potentially contains the
information on the line broadening sources (e.g., pressure broadening,
rotational broadening, winds). However, here $\sigma_{\mathrm{Na}}$ and
A$_{\mathrm{Na}}$ are degenerate and therefore we cannot robustly interpret the
physical conditions of the atmosphere directly from them. It is only the
product of these two values that indicates a physical meaning from
equivalent width of the exoplanetary sodium that blocks the star.
We note that in principle it is possible to break this degeneracy by comparing the sodium line profile to the profile obtained from the complete transmission spectrum. The complete spectrum can be obtained from the division of in-transit spectra by the out-of-transit spectra, similar to the methods of \citet{Wyttenbach2015} and \citet{Redfield2008}. However, we did not use that method because of very high intrinsic variability and activity of this star, especially during the time of this observation, which prevented us from achieving the required alignment between the spectra. In our upcoming paper, we will  investigate this further by analyzing a planet orbiting a less active star.

\subsection{Excess light-curve depths}

\begin{figure*}
\centering
\includegraphics[width=.8\textwidth]{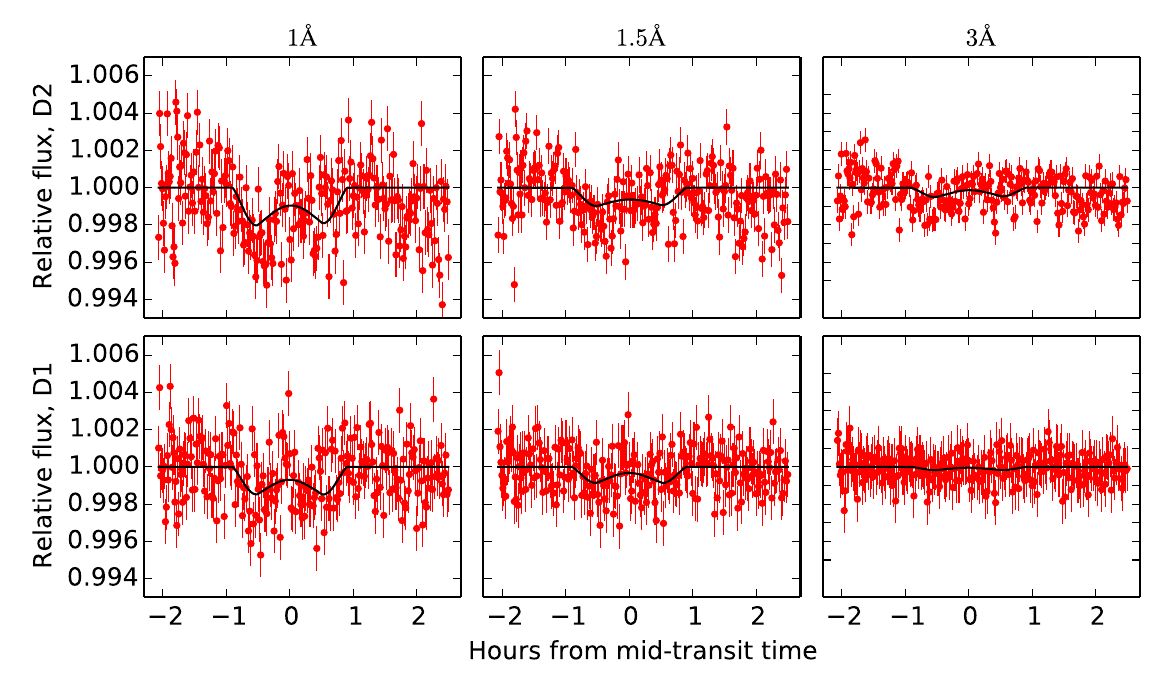}
\caption{The excess light curves in three integration passbands for both sodium D$_{1}$ and D$_{2}$ after correcting for the flare and differential limb-darkening components.}
\label{fig:neat_plot}
\end{figure*}

For a better visualization of the outcome, in Figure \ref{fig:neat_plot} we show all of the excess light curves after removing the flare and differential limb-darkening components. The only model plotted over the data in this figure is the planetary RV model component.
In our models, we did not  use any transit light-curve
model directly and so the amount of the additional depth that has been added to the
light curve  here can be achieved by measuring the
equivalent width of the exoplanetary sodium line. The additional
excess to the light curve is then the equivalent width divided by the
integrated flux of each passband divided by the reference band level.
Table~\ref{tbl:width-depth} shows the values of the excess
light curves for the three integration bands (\mbox{1 \AA}, \mbox{1.5
  \AA}, and, \mbox{3\AA}) as the average of sodium D$_{2}$ and D$_{1}$
lines. As expected, these light curves show that the depth of the
excess light curve increases with the decrease of the width of the
integration passband. This is due to a higher contribution of the star
in the wider passbands compared to that of the planet.

\begin{table}[ht!]
\centering
\caption {Computed values for the relative absorption depth in [\%] of the light curve.}
\label{tbl:width-depth}
\begin{tabular}{ l c}
  \hline\hline
  Stellar line               & Excess depth \\
  \hline
  1\AA &  \\
  Average(D$_{1}$, D$_{2}$)    & 0.72  $\pm$ 0.25 \\
  \hline
  1.5\AA &  \\
  Average(D$_{1}$, D$_{2}$)    &  0.34  $\pm$ 0.11 \\
  \hline
  3\AA &  \\
  Average(D$_{1}$, D$_{2}$)    &  0.20  $\pm$ 0.07 \\
  \hline
\end{tabular}
\end{table}

In all cases the derived excess depth obtained from the Na D$_{2}$
line for a given integration band is deeper than the
one obtained from the Na D$_{1}$ line. This has also been found in
\cite{Huitson2012}. One possible scenario is explained by looking at
the strength of the absorption signal, which is proportional to the
contrast between the stellar sodium line and the planetary sodium
line. In the case of Na D$_{2}$ the stellar line is deeper and
therefore the contrast is smaller. Thus, the signal in Na D$_{2}$
appears larger than  the transmission signal in Na D$_{1}$.

\subsection{Potassium and neutral calcium lines}

Potassium and calcium are also among the alkali metals. After sodium,
for the computed effective temperature of \mbox{HD~189733b,} models of
exoplanet atmospheres predict potassium as a prominent absorption
feature \citep{Seager2000,Fortney2010}. To investigate the presence of
these alkali metals in the \mbox{HD~189733b} atmosphere, we apply the
same method to study the potassium line at \mbox{7699 $\AA$} and a
pronounced neutral calcium line at \mbox{6122.22 $\AA$}. However, we
were not able to detect any signature of exoplanetary potassium and calcium
in their excess light curves within their noise. We attribute this
to the quality of the data rather than to the chemical composition of
the exoplanet atmosphere: the computed standard deviation of our
potassium excess light curve is \mbox{1.2 ppt}, while the amplitude of
potassium derived from models is predicted to be \mbox{0.24 ppt}
\citep{Pont2013}. Therefore, our data are not precise enough to
produce reliable results around these wavelengths. For the calcium
excess light curve, the standard deviation of the excess light curve
is \mbox{1.0 ppt}. Although the data is slightly more accurate, in
agreement with \citet{Pont2013} we did not detect any calcium
absorption. This could probably mean that the quality of the data in
both observations are not sufficient to detect the neutral
calcium. An alternative explanation could be that the calcium
abundance is not in the limit of detection.

\subsection{Comparison with previous work}

\begin{figure}[htp]
\centering
%% Use two temporary storage bins
\sbox0{\includegraphics[width = 0.52\textwidth]{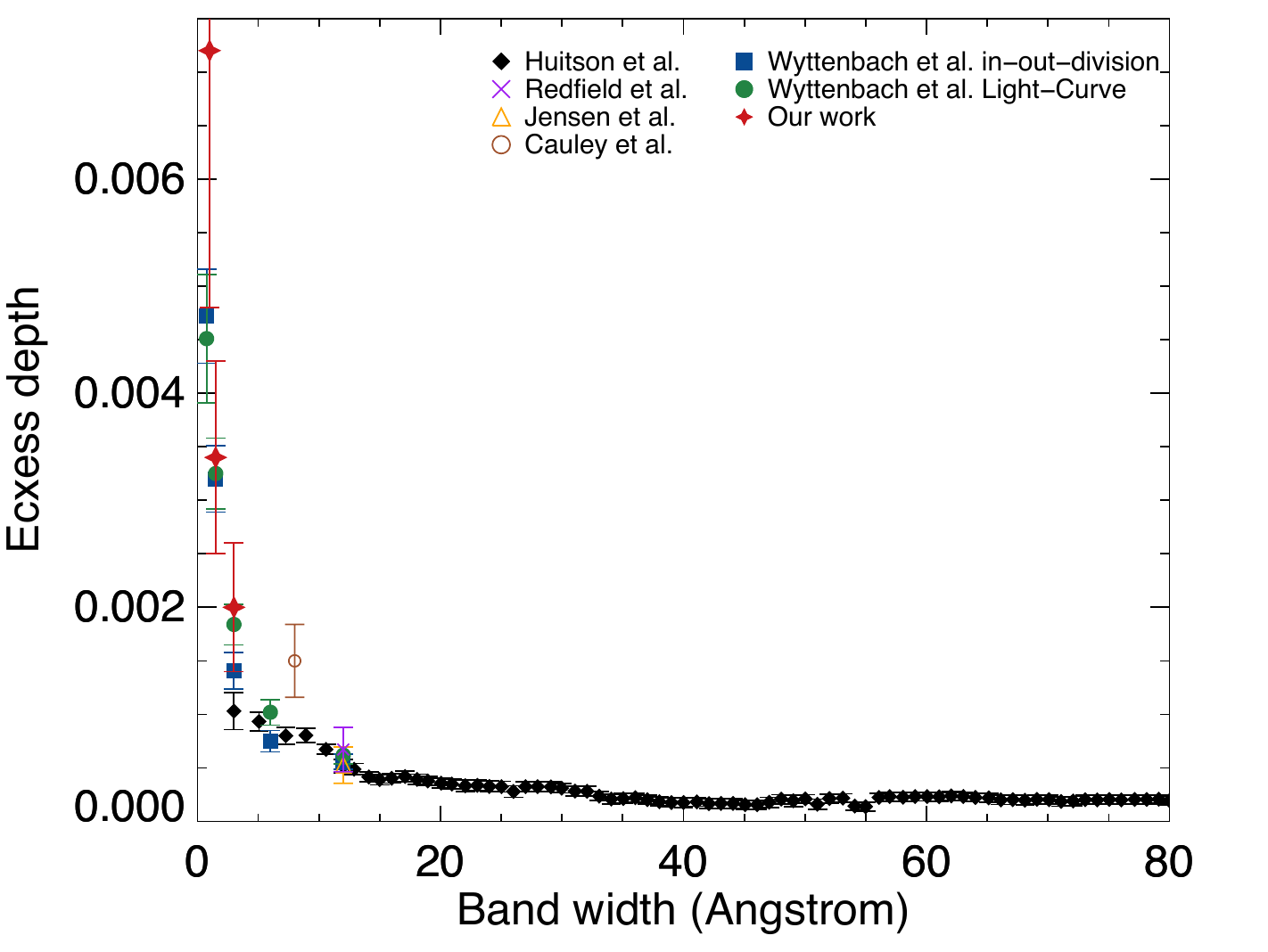}}
\sbox2{\includegraphics[scale = 0.25]{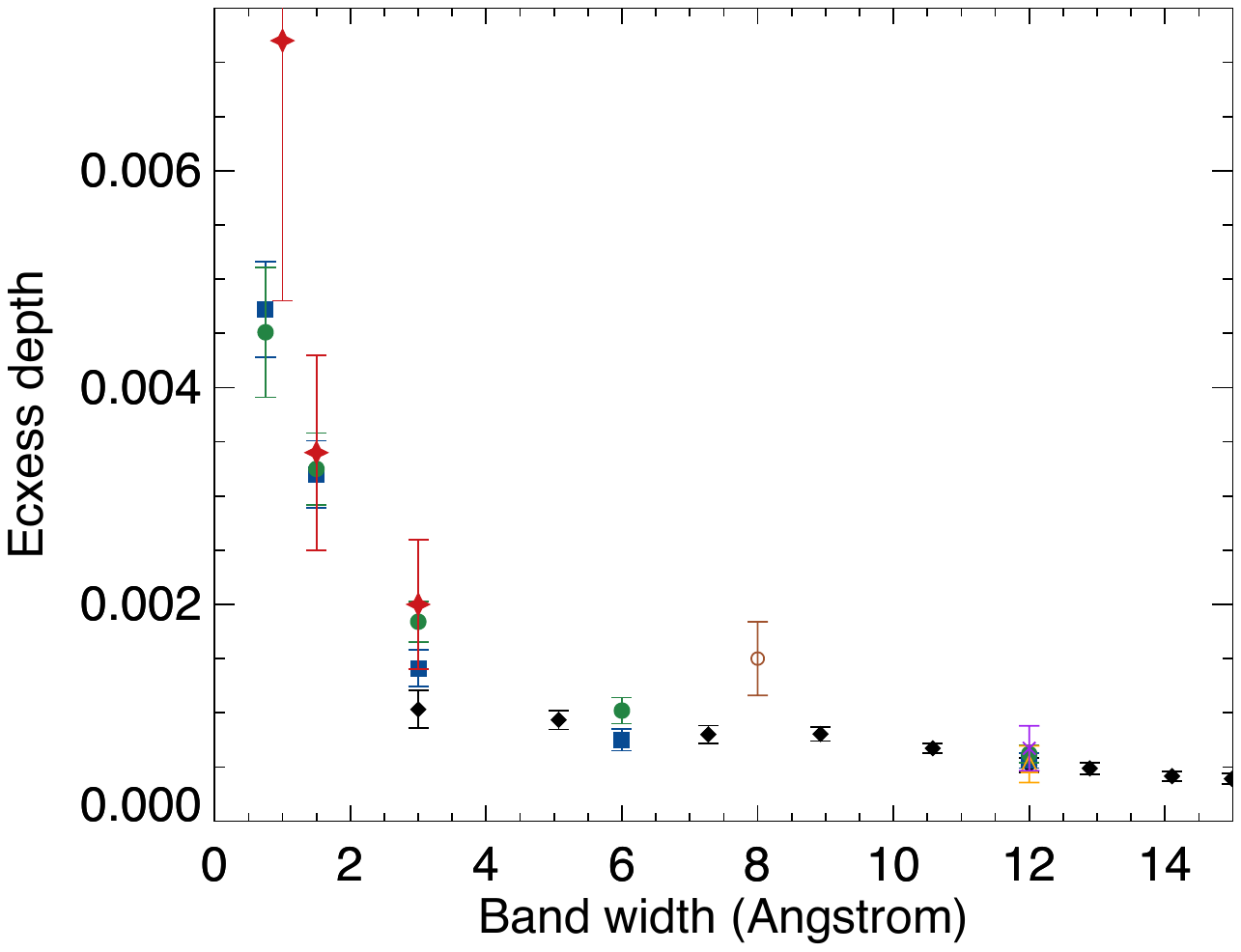}} % add the necessary scaling options
\begin{picture}(\wd0,\ht0)
\put(0,0){\usebox0}
%% Adjust the -0.7cm and -0.5cm shifts
\put(\wd0 - \wd2 - 0.9cm,\ht0 - \ht2 - 1.2cm){\usebox{2}}
\end{picture}
\caption{\label{fig:Huitson} Our averaged excess depths for each integration band centered at
  the sodium lines (red filled stars) compared to previous works.}
\end{figure}

Previous measurements of the excess absorption of the sodium in
\mbox{HD~189733b} for passbands between 1 and \mbox{80 \AA}, along
with our results, are shown in Figure~\ref{fig:Huitson}. Bibliographic
values are taken from \cite{Wyttenbach2015}, \citet{Huitson2012},
\cite{Redfield2008}, \cite{Jensen2011}, and \cite{Cauley2016}. 
\cite{Huitson2012} used low-resolution spectra of STIS/HST and obtained the excess depths
for wide ranges of integration bands from 3 to \mbox{80 \AA} around
the sodium feature (black filled diamonds). \cite{Wyttenbach2015} used
high-resolution data obtained with HARPS and performed ground-based
transmission spectroscopy using integration bands of 0.75, 1.5, 3, 6,
and \mbox{12 \AA}. They used two approaches: one by fitting a transit
model on the excess light curve (green filled circles) and the other 
by investigating the residuals of the in-transit divided by the out-of-transit
spectra (in-out division;  blue filled squares).
\cite{Redfield2008} and \cite{Jensen2011} both used the same set of data taken by the high-resolution spectrograph mounted at the Hobby Eberly Telescope (HET)
and tried to detect the exo-atmosphere by applying the in-out division approach (yellow unfilled triangle and purple cross). Finally, using the same approach, \cite{Cauley2016} analyzed a single transit observation of HiRES on Keck (unfilled brown circle).
In our work we introduced a new approach for an accurate
measurement of the exoplanetary transmission signal. Compared to previous measurements this approach is unbiased by stellar differential limb-darkening effect, stellar flaring activity, and the changing RV bump. Our estimated values of the
additional absorption by the exoplanetary atmospheric sodium are indicated by the red filled stars in Figure \ref{fig:Huitson}. 
In comparison to other studies, our data points show stronger
absorption with larger error bars. The main reason for this difference must be the correction of the bulgy shape in the middle of the transit excess light curve caused by differential limb-darkening and changing radial velocity effects. In other words, not considering an increase in flux near the mid-transit time would underestimate the absorption depth, thus producing smaller values.
The  error bars in our work are larger by about a factor of 3. This is reasonable since our model has more components and therefore consists of fewer degrees of freedom.

The differential limb-darkening model calculations performed in this
work can be compared with those computed by \citet{Czesla2015}. The
purpose of this paper  is to consider the line cores to actually
search for the planet signatures. Therefore, the cores of the lines
are investigated, where the effects of the flare and the planets are
present. \citet{Czesla2015}  only considered the wings of the sodium lines where the effects of stellar activity and the planet are
not present  (or only marginally). Furthermore, we use ratios rather than
differences, but as shown in \citeauthor{Czesla2015}, this is a small
effect for small-amplitude variability (Eq. 7 in \citealt{Czesla2015}).
Moreover, \citet{Czesla2015} use Kurucz models, while we use PHOENIX
angle resolved synthetic spectra. A difference between PHOENIX and
Kurucz is that the former is calculated in spherical symmetry, the latter in 
plane parallel, which is justified because the
atmosphere in main sequence stars is thin and the curvature does not play
a role. However, getting the LD correct for our needed order of
precision, the plane parallel geometry is probably insufficient. In
this respect, the phoenix model is considered to be better.

\subsection{Rough estimation of Na number density in the \mbox{HD~189733b} atmosphere}

Stellar abundances can be obtained from equivalent widths (EWs) of
spectral lines, using a relation called the ``curve of growth''. For
small equivalent widths (i.e., weak lines) the equivalent width of a
line is proportional to its number density. In this work we computed the
equivalent width of \mbox{HD~189733b} as the product between the line
depth and the line width, ln(EW) = ln(A$_{\mathrm{Na}}\times$FWHM) =
ln(A$_{\mathrm{Na}}\times$2$\sqrt{2 ln(2)}~\sigma_{\mathrm{Na}}$) = -5.02, corresponding
as expected to the weak line regime. Assuming a scale height of 100 km
above the clouds, the number density of sodium in the atmosphere of
\mbox{HD~189733b} is 10$^4$ atoms/cm$^3$, close to the value
that \citet{Heng2015} estimated.

\subsection{Further considerations in the model budget}

Despite carrying out  a meticulous analysis of the model
components in this work, there are other aspects that might have a second-order
impact on our data. For example, stellar spots and stellar rotation
are two physical effects that deform the stellar spectral line shape
and, therefore, might influence the derived excess light curves. In
this work these effects were ignored for  the  two reasons. First, 
with respect to stellar rotation, \mbox{HD~189733} is a slow rotating
star \citep[$\sim$12 days]{Winn2007} and since the exposures are
equally distributed along the transit this effect, if any, should be
averaged out in our all-transit integrations
\citep{Dravins2015,Wyttenbach2015}. Second,  to
measure the influence of occulted and unocculted stellar spots, in addition to the
rotational period and amplitude of the modulation of the total stellar
flux (which would give us an idea of the spot coverage at the moment
of the observations) we would have needed simultaneous broad-band
observations \citep{Pont2013,McCullough2014}. Indeed, the use of
simultaneous observations with low-resolution instruments is a good
way to detect spot crossing events. A question that we will be able to  answer 
with new high-resolution observations simultaneous to broad-band
data is whether the change in excess depth observed in this work
could have been created by unaccounted spot signal. With only the present data set
we cannot assess this.

%-------------------------------------------------------------------
%                               CONCLUSION
%-------------------------------------------------------------------

\section{Conclusions}
\label{sec:conclusions}

We used high-resolution spectra of UVES to measure the sodium
absorption in the transmission spectrum of the hot Jupiter
\mbox{HD~189733b}. In this work we applied a new approach based on the
changing radial velocity of the exoplanet.  To this end, we analyzed
six excess light curves in sodium D$_1$ and D$_2$ that were integrated
in three different bands. We modeled a combination of three main
effects on each data set and extracted some information about the
exo-atmospheric sodium. These effects are a stellar flare starting
close to mid-transit time, the differential limb-darkening
effects, and the planetary sodium line profile that is moving inside
the stellar sodium line because of the change in the radial velocity
of its orbit.

We confirm the ground-based detection of sodium in the
upper atmosphere of \mbox{HD~189733b}. Close to the mid-transit time a
bump appears in the excess light curves, which is caused by the radial
velocity changes of the planet while moving in its orbit with respect
to its parent star. By modeling this effect we estimate the equivalent
width of the exoplanetary sodium line by an estimation of
its width and depth. In the future, with the improvement in the quality of
observations, this might be used to better constrain the physical
conditions in the upper atmosphere of exoplanets as part of the
atmospheric models. The average equivalent width of
the two sodium lines is 0.0023 $\pm$ 0.0010 $\AA$.  The average depth of the excess
light curve is \mbox{0.72 $\pm$ 0.25 \%} at \mbox{1 \AA},
\mbox{0.34$\pm$ 0.11 \%} at \mbox{1.5 \AA}, and \mbox{0.20 $\pm$
  0.06 \%} at \mbox{3 \AA}, in good agreement with previous research.
 
Finally,  in this work we learn the following. First, in high-resolution
transmission spectroscopy at any wavelength, Ca H \& K or H$_{\alpha}$
should be always observed simultaneously when monitoring flaring
activity consistency. Second, depending on some parameters, such as the
temperature of the star, the width of the stellar sodium line, and the
orbital velocity of the planet, the bump that appears in the mid-point
of the excess light curves can originate from changing radial
velocity effects and/or from the differential limb-darkening effect. A
careful treatment of these two factors must be carried out. Finally, in an
observation with short cadence and a large number of data points, the
radial velocity bump can be better studied.

\begin{acknowledgements} 

S. Khalafinejad acknowledges funding by the DFG in the framework of
RTG 1351 and thanks I. Snellen, S. Albrecht, and the anonymous referee for their suggestions
and advice. Accomplishment of this work was obtained through
assistance, advice, and scientific discussions with many other people,
especially at Hamburg Observatory and Center for Astrophysics
(CfA). S. Khalafinejad would like to specifically thank M. Holman, S. Czesla, M. G\"udel,
B. Fuhrmeister, A. K. Dupree, M. Payne, H. M\"uller, P. Ioannidis, and
M. Salz for their valuable support and assistance. C. von
Essen acknowledges funding for the Stellar Astrophysics Centre
provided by The Danish National Research Foundation (grant
No. DNRF106). T. Klocov\'a acknowledges support from RTG 1351 and DFG
project CZ 222/1-1.

\end{acknowledgements}

\bibliography{2016-09-04_HD189_submit}

\begin{thebibliography}{65}
\expandafter\ifx\csname natexlab\endcsname\relax\def\natexlab#1{#1}\fi

\bibitem[{{Agol} {et~al.}(2010){Agol}, {Cowan}, {Knutson}, {Deming}, {Steffen},
  {Henry}, \& {Charbonneau}}]{Agol2010}
{Agol}, E., {Cowan}, N.~B., {Knutson}, H.~A., {et~al.} 2010, \apj, 721, 1861

\bibitem[{{Albrecht}(2008)}]{Albrecht2008Thes}
{Albrecht}, S. 2008, PhD thesis, Leiden Observatory, Leiden University,
  P.O.~Box 9513, 2300 RA Leiden, The Netherlands

\bibitem[{{Birkby} {et~al.}(2013){Birkby}, {de Kok}, {Brogi}, {de Mooij},
  {Schwarz}, {Albrecht}, \& {Snellen}}]{Birkby2013}
{Birkby}, J.~L., {de Kok}, R.~J., {Brogi}, M., {et~al.} 2013, \mnras, 436, L35

\bibitem[{{Borucki} {et~al.}(2010){Borucki}, {Koch}, {Basri}, {Batalha},
  {Brown}, {Caldwell}, {Caldwell}, {Christensen-Dalsgaard}, {Cochran},
  {DeVore}, {Dunham}, {Dupree}, {Gautier}, {Geary}, {Gilliland}, {Gould},
  {Howell}, {Jenkins}, {Kondo}, {Latham}, {Marcy}, {Meibom}, {Kjeldsen},
  {Lissauer}, {Monet}, {Morrison}, {Sasselov}, {Tarter}, {Boss}, {Brownlee},
  {Owen}, {Buzasi}, {Charbonneau}, {Doyle}, {Fortney}, {Ford}, {Holman},
  {Seager}, {Steffen}, {Welsh}, {Rowe}, {Anderson}, {Buchhave}, {Ciardi},
  {Walkowicz}, {Sherry}, {Horch}, {Isaacson}, {Everett}, {Fischer}, {Torres},
  {Johnson}, {Endl}, {MacQueen}, {Bryson}, {Dotson}, {Haas}, {Kolodziejczak},
  {Van Cleve}, {Chandrasekaran}, {Twicken}, {Quintana}, {Clarke}, {Allen},
  {Li}, {Wu}, {Tenenbaum}, {Verner}, {Bruhweiler}, {Barnes}, \&
  {Prsa}}]{Borucki2010}
{Borucki}, W.~J., {Koch}, D., {Basri}, G., {et~al.} 2010, Science, 327, 977

\bibitem[{{Boyajian} {et~al.}(2015){Boyajian}, {von Braun}, {Feiden}, {Huber},
  {Basu}, {Demarque}, {Fischer}, {Schaefer}, {Mann}, {White}, {Maestro},
  {Brewer}, {Lamell}, {Spada}, {L{\'o}pez-Morales}, {Ireland}, {Farrington},
  {van Belle}, {Kane}, {Jones}, {ten Brummelaar}, {Ciardi}, {McAlister},
  {Ridgway}, {Goldfinger}, {Turner}, \& {Sturmann}}]{Boyajian2015}
{Boyajian}, T., {von Braun}, K., {Feiden}, G.~A., {et~al.} 2015, \mnras, 447,
  846

\bibitem[{{Brogi} {et~al.}(2012){Brogi}, {Snellen}, {de Kok}, {Albrecht},
  {Birkby}, \& {de Mooij}}]{Brogi2012}
{Brogi}, M., {Snellen}, I.~A.~G., {de Kok}, R.~J., {et~al.} 2012, \nat, 486,
  502

\bibitem[{{Brown}(2001)}]{Brown2001}
{Brown}, T.~M. 2001, \apj, 553, 1006

\bibitem[{{Burrows}(2014)}]{Burrows2014}
{Burrows}, A.~S. 2014, \nat, 513, 345

\bibitem[{{Cauley} {et~al.}(2016){Cauley}, {Redfield}, {Jensen}, \&
  {Barman}}]{Cauley2016}
{Cauley}, P.~W., {Redfield}, S., {Jensen}, A.~G., \& {Barman}, T. 2016, \aj,
  152, 20

\bibitem[{{Cessateur} {et~al.}(2010){Cessateur}, {Kretzschmar}, {Dudok de Wit},
  \& {Boumier}}]{Cessateur2010}
{Cessateur}, G., {Kretzschmar}, M., {Dudok de Wit}, T., \& {Boumier}, P. 2010,
  \solphys, 263, 153

\bibitem[{{Charbonneau} {et~al.}(2000){Charbonneau}, {Brown}, {Latham}, \&
  {Mayor}}]{Charbonneau2000}
{Charbonneau}, D., {Brown}, T.~M., {Latham}, D.~W., \& {Mayor}, M. 2000, \apjl,
  529, L45

\bibitem[{{Charbonneau} {et~al.}(2002){Charbonneau}, {Brown}, {Noyes}, \&
  {Gilliland}}]{Charbonneau2002}
{Charbonneau}, D., {Brown}, T.~M., {Noyes}, R.~W., \& {Gilliland}, R.~L. 2002,
  \apj, 568, 377

\bibitem[{{Crossfield} {et~al.}(2011){Crossfield}, {Barman}, \&
  {Hansen}}]{Crossfield2011}
{Crossfield}, I.~J.~M., {Barman}, T., \& {Hansen}, B.~M.~S. 2011, \apj, 736,
  132

\bibitem[{{Czesla} {et~al.}(2015){Czesla}, {Klocov{\'a}}, {Khalafinejad},
  {Wolter}, \& {Schmitt}}]{Czesla2015}
{Czesla}, S., {Klocov{\'a}}, T., {Khalafinejad}, S., {Wolter}, U., \&
  {Schmitt}, J.~H.~M.~M. 2015, \aap, 582, A51

\bibitem[{{de Kok} {et~al.}(2013){de Kok}, {Brogi}, {Snellen}, {Birkby},
  {Albrecht}, \& {de Mooij}}]{deKok2013}
{de Kok}, R.~J., {Brogi}, M., {Snellen}, I.~A.~G., {et~al.} 2013, \aap, 554,
  A82

\bibitem[{{Dekker} {et~al.}(2000){Dekker}, {D'Odorico}, {Kaufer}, {Delabre}, \&
  {Kotzlowski}}]{Dekker2000}
{Dekker}, H., {D'Odorico}, S., {Kaufer}, A., {Delabre}, B., \& {Kotzlowski}, H.
  2000, in Society of Photo-Optical Instrumentation Engineers (SPIE) Conference
  Series, Vol. 4008, Optical and IR Telescope Instrumentation and Detectors,
  ed. M.~{Iye} \& A.~F. {Moorwood}, 534--545

\bibitem[{{Deming} {et~al.}(2013){Deming}, {Wilkins}, {McCullough}, {Burrows},
  {Fortney}, {Agol}, {Dobbs-Dixon}, {Madhusudhan}, {Crouzet}, {Desert},
  {Gilliland}, {Haynes}, {Knutson}, {Line}, {Magic}, {Mandell}, {Ranjan},
  {Charbonneau}, {Clampin}, {Seager}, \& {Showman}}]{Deming2013}
{Deming}, D., {Wilkins}, A., {McCullough}, P., {et~al.} 2013, \apj, 774, 95

\bibitem[{{D{\'e}sert} {et~al.}(2009){D{\'e}sert}, {Lecavelier des Etangs},
  {H{\'e}brard}, {Sing}, {Ehrenreich}, {Ferlet}, \&
  {Vidal-Madjar}}]{Desert2009}
{D{\'e}sert}, J.-M., {Lecavelier des Etangs}, A., {H{\'e}brard}, G., {et~al.}
  2009, \apj, 699, 478

\bibitem[{{D{\'e}sert} {et~al.}(2011){D{\'e}sert}, {Sing}, {Vidal-Madjar},
  {H{\'e}brard}, {Ehrenreich}, {Lecavelier Des Etangs}, {Parmentier}, {Ferlet},
  \& {Henry}}]{Desert2011}
{D{\'e}sert}, J.-M., {Sing}, D., {Vidal-Madjar}, A., {et~al.} 2011, \aap, 526,
  A12

\bibitem[{{Dravins} {et~al.}(2015){Dravins}, {Ludwig}, {Dahlen}, \&
  {Pazira}}]{Dravins2015}
{Dravins}, D., {Ludwig}, H.-G., {Dahlen}, E., \& {Pazira}, H. 2015, in
  Cambridge Workshop on Cool Stars, Stellar Systems, and the Sun, Vol.~18, 18th
  Cambridge Workshop on Cool Stars, Stellar Systems, and the Sun, ed. G.~T.
  {van Belle} \& H.~C. {Harris}, 853--868

\bibitem[{{Ehrenreich} {et~al.}(2015){Ehrenreich}, {Bourrier}, {Wheatley},
  {Lecavelier des Etangs}, {H{\'e}brard}, {Udry}, {Bonfils}, {Delfosse},
  {D{\'e}sert}, {Sing}, \& {Vidal-Madjar}}]{Ehrenreich2015}
{Ehrenreich}, D., {Bourrier}, V., {Wheatley}, P.~J., {et~al.} 2015, \nat, 522,
  459

\bibitem[{{Foreman-Mackey} {et~al.}(2013){Foreman-Mackey}, {Conley},
  {Meierjurgen Farr}, {Hogg}, {Long}, {Marshall}, {Price-Whelan}, {Sanders}, \&
  {Zuntz}}]{Foreman2013}
{Foreman-Mackey}, D., {Conley}, A., {Meierjurgen Farr}, W., {et~al.} 2013,
  {emcee: The MCMC Hammer}, Astrophysics Source Code Library

\bibitem[{{Fortney} {et~al.}(2010){Fortney}, {Shabram}, {Showman}, {Lian},
  {Freedman}, {Marley}, \& {Lewis}}]{Fortney2010}
{Fortney}, J.~J., {Shabram}, M., {Showman}, A.~P., {et~al.} 2010, \apj, 709,
  1396

\bibitem[{{Gibson} {et~al.}(2013){Gibson}, {Aigrain}, {Barstow}, {Evans},
  {Fletcher}, \& {Irwin}}]{Gibson2013}
{Gibson}, N.~P., {Aigrain}, S., {Barstow}, J.~K., {et~al.} 2013, \mnras, 428,
  3680

\bibitem[{{Hartman} {et~al.}(2004){Hartman}, {Bakos}, {Stanek}, \&
  {Noyes}}]{Hartman2004}
{Hartman}, J.~D., {Bakos}, G., {Stanek}, K.~Z., \& {Noyes}, R.~W. 2004, \aj,
  128, 1761

\bibitem[{{Hauschildt} \& {Baron}(1999)}]{Hauschildt1999}
{Hauschildt}, P.~H. \& {Baron}, E. 1999, Journal of Computational and Applied
  Mathematics, 109, 41

\bibitem[{{Heng} {et~al.}(2015){Heng}, {Wyttenbach}, {Lavie}, {Sing},
  {Ehrenreich}, \& {Lovis}}]{Heng2015}
{Heng}, K., {Wyttenbach}, A., {Lavie}, B., {et~al.} 2015, \apjl, 803, L9

\bibitem[{{Henry} \& {Winn}(2008)}]{Henry2008}
{Henry}, G.~W. \& {Winn}, J.~N. 2008, \aj, 135, 68

\bibitem[{{Hoeijmakers} {et~al.}(2015){Hoeijmakers}, {de Kok}, {Snellen},
  {Brogi}, {Birkby}, \& {Schwarz}}]{Hoeijmakers2015}
{Hoeijmakers}, H.~J., {de Kok}, R.~J., {Snellen}, I.~A.~G., {et~al.} 2015,
  \aap, 575, A20

\bibitem[{{Huitson} {et~al.}(2012){Huitson}, {Sing}, {Vidal-Madjar},
  {Ballester}, {Lecavelier des Etangs}, {D{\'e}sert}, \& {Pont}}]{Huitson2012}
{Huitson}, C.~M., {Sing}, D.~K., {Vidal-Madjar}, A., {et~al.} 2012, \mnras,
  422, 2477

\bibitem[{{Husser} {et~al.}(2015){Husser}, {Kamann}, \&
  {Dreizler}}]{Husser2015}
{Husser}, T.-O., {Kamann}, S., \& {Dreizler}, S. 2015, \aap\ submitted

\bibitem[{{Husser} \& {Ulbrich}(2014)}]{Husser2014}
{Husser}, T.-O. \& {Ulbrich}, K. 2014, in Astronomical Society of India
  Conference Series, Vol.~11, Astronomical Society of India Conference Series,
  53--56

\bibitem[{{Husser} {et~al.}(2013){Husser}, {Wende-von Berg}, {Dreizler},
  {Homeier}, {Reiners}, {Barman}, \& {Hauschildt}}]{Husser2013}
{Husser}, T.-O., {Wende-von Berg}, S., {Dreizler}, S., {et~al.} 2013, \aap,
  553, A6

\bibitem[{{Jensen} {et~al.}(2011){Jensen}, {Redfield}, {Endl}, {Cochran},
  {Koesterke}, \& {Barman}}]{Jensen2011}
{Jensen}, A.~G., {Redfield}, S., {Endl}, M., {et~al.} 2011, \apj, 743, 203

\bibitem[{{Koch} {et~al.}(2010){Koch}, {Borucki}, {Basri}, {Batalha}, {Brown},
  {Caldwell}, {Christensen-Dalsgaard}, {Cochran}, {DeVore}, {Dunham},
  {Gautier}, {Geary}, {Gilliland}, {Gould}, {Jenkins}, {Kondo}, {Latham},
  {Lissauer}, {Marcy}, {Monet}, {Sasselov}, {Boss}, {Brownlee}, {Caldwell},
  {Dupree}, {Howell}, {Kjeldsen}, {Meibom}, {Morrison}, {Owen}, {Reitsema},
  {Tarter}, {Bryson}, {Dotson}, {Gazis}, {Haas}, {Kolodziejczak}, {Rowe}, {Van
  Cleve}, {Allen}, {Chandrasekaran}, {Clarke}, {Li}, {Quintana}, {Tenenbaum},
  {Twicken}, \& {Wu}}]{Koch2010}
{Koch}, D.~G., {Borucki}, W.~J., {Basri}, G., {et~al.} 2010, \apjl, 713, L79

\bibitem[{{Kreidberg} {et~al.}(2014){Kreidberg}, {Bean}, {D{\'e}sert},
  {Benneke}, {Deming}, {Stevenson}, {Seager}, {Berta-Thompson}, {Seifahrt}, \&
  {Homeier}}]{Kreidberg2014}
{Kreidberg}, L., {Bean}, J.~L., {D{\'e}sert}, J.-M., {et~al.} 2014, \nat, 505,
  69

\bibitem[{{Kulow} {et~al.}(2014){Kulow}, {France}, {Linsky}, \&
  {Loyd}}]{Kulow2014}
{Kulow}, J.~R., {France}, K., {Linsky}, J., \& {Loyd}, R.~O.~P. 2014, \apj,
  786, 132

\bibitem[{{Lockwood} {et~al.}(2014){Lockwood}, {Johnson}, {Bender}, {Carr},
  {Barman}, {Richert}, \& {Blake}}]{Lockwood2014}
{Lockwood}, A.~C., {Johnson}, J.~A., {Bender}, C.~F., {et~al.} 2014, \apjl,
  783, L29

\bibitem[{{Louden} \& {Wheatley}(2015)}]{Louden2015}
{Louden}, T. \& {Wheatley}, P.~J. 2015, \apjl, 814, L24

\bibitem[{{Mandel} \& {Agol}(2002)}]{Agol2002}
{Mandel}, K. \& {Agol}, E. 2002, \apjl, 580, L171

\bibitem[{{McCullough} {et~al.}(2014){McCullough}, {Crouzet}, {Deming}, \&
  {Madhusudhan}}]{McCullough2014}
{McCullough}, P.~R., {Crouzet}, N., {Deming}, D., \& {Madhusudhan}, N. 2014,
  \apj, 791, 55

\bibitem[{{Nikolov} {et~al.}(2015){Nikolov}, {Sing}, {Burrows}, {Fortney},
  {Henry}, {Pont}, {Ballester}, {Aigrain}, {Wilson}, {Huitson}, {Gibson},
  {D{\'e}sert}, {Etangs}, {Showman}, {Vidal-Madjar}, {Wakeford}, \&
  {Zahnle}}]{Nikolov2015}
{Nikolov}, N., {Sing}, D.~K., {Burrows}, A.~S., {et~al.} 2015, \mnras, 447, 463

\bibitem[{{Pollacco} {et~al.}(2006){Pollacco}, {Skillen}, {Collier Cameron},
  {Christian}, {Hellier}, {Irwin}, {Lister}, {Street}, {West}, {Anderson},
  {Clarkson}, {Deeg}, {Enoch}, {Evans}, {Fitzsimmons}, {Haswell}, {Hodgkin},
  {Horne}, {Kane}, {Keenan}, {Maxted}, {Norton}, {Osborne}, {Parley}, {Ryans},
  {Smalley}, {Wheatley}, \& {Wilson}}]{Pollacco2006}
{Pollacco}, D.~L., {Skillen}, I., {Collier Cameron}, A., {et~al.} 2006, \pasp,
  118, 1407

\bibitem[{{Pont} {et~al.}(2008){Pont}, {Knutson}, {Gilliland}, {Moutou}, \&
  {Charbonneau}}]{Pont2008}
{Pont}, F., {Knutson}, H., {Gilliland}, R.~L., {Moutou}, C., \& {Charbonneau},
  D. 2008, \mnras, 385, 109

\bibitem[{{Pont} {et~al.}(2013){Pont}, {Sing}, {Gibson}, {Aigrain}, {Henry}, \&
  {Husnoo}}]{Pont2013}
{Pont}, F., {Sing}, D.~K., {Gibson}, N.~P., {et~al.} 2013, \mnras, 432, 2917

\bibitem[{{Poppenhaeger} {et~al.}(2013){Poppenhaeger}, {Schmitt}, \&
  {Wolk}}]{Poppenhaeger2013}
{Poppenhaeger}, K., {Schmitt}, J.~H.~M.~M., \& {Wolk}, S.~J. 2013, \apj, 773,
  62

\bibitem[{{Redfield} {et~al.}(2008){Redfield}, {Endl}, {Cochran}, \&
  {Koesterke}}]{Redfield2008}
{Redfield}, S., {Endl}, M., {Cochran}, W.~D., \& {Koesterke}, L. 2008, \apjl,
  673, L87

\bibitem[{{Rodler} {et~al.}(2012){Rodler}, {Lopez-Morales}, \&
  {Ribas}}]{Rodler2012}
{Rodler}, F., {Lopez-Morales}, M., \& {Ribas}, I. 2012, \apjl, 753, L25

\bibitem[{{Schwarz}(1978)}]{Schwarz1978}
{Schwarz}, G. 1978, The annals of statistics, 6, 461

\bibitem[{{Seager} \& {Sasselov}(2000)}]{Seager2000}
{Seager}, S. \& {Sasselov}, D.~D. 2000, \apj, 537, 916

\bibitem[{{Sing} {et~al.}(2016){Sing}, {Fortney}, {Nikolov}, {Wakeford},
  {Kataria}, {Evans}, {Aigrain}, {Ballester}, {Burrows}, {Deming},
  {D{\'e}sert}, {Gibson}, {Henry}, {Huitson}, {Knutson}, {Etangs}, {Pont},
  {Showman}, {Vidal-Madjar}, {Williamson}, \& {Wilson}}]{Sing2016}
{Sing}, D.~K., {Fortney}, J.~J., {Nikolov}, N., {et~al.} 2016, \nat, 529, 59

\bibitem[{{Sing} {et~al.}(2012){Sing}, {Huitson}, {Lopez-Morales}, {Pont},
  {D{\'e}sert}, {Ehrenreich}, {Wilson}, {Ballester}, {Fortney}, {Lecavelier des
  Etangs}, \& {Vidal-Madjar}}]{Sing2012}
{Sing}, D.~K., {Huitson}, C.~M., {Lopez-Morales}, M., {et~al.} 2012, \mnras,
  426, 1663

\bibitem[{{Sing} {et~al.}(2011){Sing}, {Pont}, {Aigrain}, {Charbonneau},
  {D{\'e}sert}, {Gibson}, {Gilliland}, {Hayek}, {Henry}, {Knutson}, {Lecavelier
  Des Etangs}, {Mazeh}, \& {Shporer}}]{Sing2011}
{Sing}, D.~K., {Pont}, F., {Aigrain}, S., {et~al.} 2011, \mnras, 416, 1443

\bibitem[{{Sing} {et~al.}(2008){Sing}, {Vidal-Madjar}, {D{\'e}sert},
  {Lecavelier des Etangs}, \& {Ballester}}]{Sing2008}
{Sing}, D.~K., {Vidal-Madjar}, A., {D{\'e}sert}, J.-M., {Lecavelier des
  Etangs}, A., \& {Ballester}, G. 2008, \apj, 686, 658

\bibitem[{{Snellen} {et~al.}(2015){Snellen}, {de Kok}, {Birkby}, {Brandl},
  {Brogi}, {Keller}, {Kenworthy}, {Schwarz}, \& {Stuik}}]{Snellen2015}
{Snellen}, I., {de Kok}, R., {Birkby}, J.~L., {et~al.} 2015, \aap, 576, A59

\bibitem[{{Snellen} {et~al.}(2008){Snellen}, {Albrecht}, {de Mooij}, \& {Le
  Poole}}]{Snellen2008}
{Snellen}, I.~A.~G., {Albrecht}, S., {de Mooij}, E.~J.~W., \& {Le Poole}, R.~S.
  2008, \aap, 487, 357

\bibitem[{{Snellen} {et~al.}(2010){Snellen}, {de Kok}, {de Mooij}, \&
  {Albrecht}}]{Snellen2010}
{Snellen}, I.~A.~G., {de Kok}, R.~J., {de Mooij}, E.~J.~W., \& {Albrecht}, S.
  2010, \nat, 465, 1049

\bibitem[{{Tinetti} {et~al.}(2007){Tinetti}, {Carey}, {Allard}, {Ballester},
  {Barber}, {Beaulieu}, {Liang}, {Noriega-Crespo}, {Ribas}, {Selsis}, {Sing},
  {Tennyson}, \& {Yung}}]{Tinetti2007}
{Tinetti}, G., {Carey}, S., {Allard}, N., {et~al.} 2007, Spitzer Proposal, 461

\bibitem[{{Torres} {et~al.}(2008){Torres}, {Winn}, \& {Holman}}]{Torres2008}
{Torres}, G., {Winn}, J.~N., \& {Holman}, M.~J. 2008, \apj, 677, 1324

\bibitem[{{Vidal-Madjar} {et~al.}(2011){Vidal-Madjar}, {Sing}, {Lecavelier Des
  Etangs}, {Ferlet}, {D{\'e}sert}, {H{\'e}brard}, {Boisse}, {Ehrenreich}, \&
  {Moutou}}]{Vidal-Madjar2011}
{Vidal-Madjar}, A., {Sing}, D.~K., {Lecavelier Des Etangs}, A., {et~al.} 2011,
  \aap, 527, A110

\bibitem[{{von Essen} {et~al.}(2013){von Essen}, {Schr{\"o}ter}, {Agol}, \&
  {Schmitt}}]{vonEssen2013}
{von Essen}, C., {Schr{\"o}ter}, S., {Agol}, E., \& {Schmitt}, J.~H.~M.~M.
  2013, \aap, 555, A92

\bibitem[{{Winn} {et~al.}(2007){Winn}, {Holman}, {Henry}, {Roussanova}, {Enya},
  {Yoshii}, {Shporer}, {Mazeh}, {Johnson}, {Narita}, \& {Suto}}]{Winn2007}
{Winn}, J.~N., {Holman}, M.~J., {Henry}, G.~W., {et~al.} 2007, \aj, 133, 1828

\bibitem[{{Wood} {et~al.}(2011){Wood}, {Maxted}, {Smalley}, \&
  {Iro}}]{Wood2011}
{Wood}, P.~L., {Maxted}, P.~F.~L., {Smalley}, B., \& {Iro}, N. 2011, \mnras,
  412, 2376

\bibitem[{{Wyttenbach} {et~al.}(2015){Wyttenbach}, {Ehrenreich}, {Lovis},
  {Udry}, \& {Pepe}}]{Wyttenbach2015}
{Wyttenbach}, A., {Ehrenreich}, D., {Lovis}, C., {Udry}, S., \& {Pepe}, F.
  2015, \aap, 577, A62

\bibitem[{{Zhou} \& {Bayliss}(2012)}]{Zhou2012}
{Zhou}, G. \& {Bayliss}, D.~D.~R. 2012, \mnras, 426, 2483

\end{thebibliography}

\newpage

\setcounter{figure}{0} \renewcommand{\thefigure}{A.\arabic{figure}}

\begin{figure*}
\centering
\includegraphics[width=0.8\textwidth]{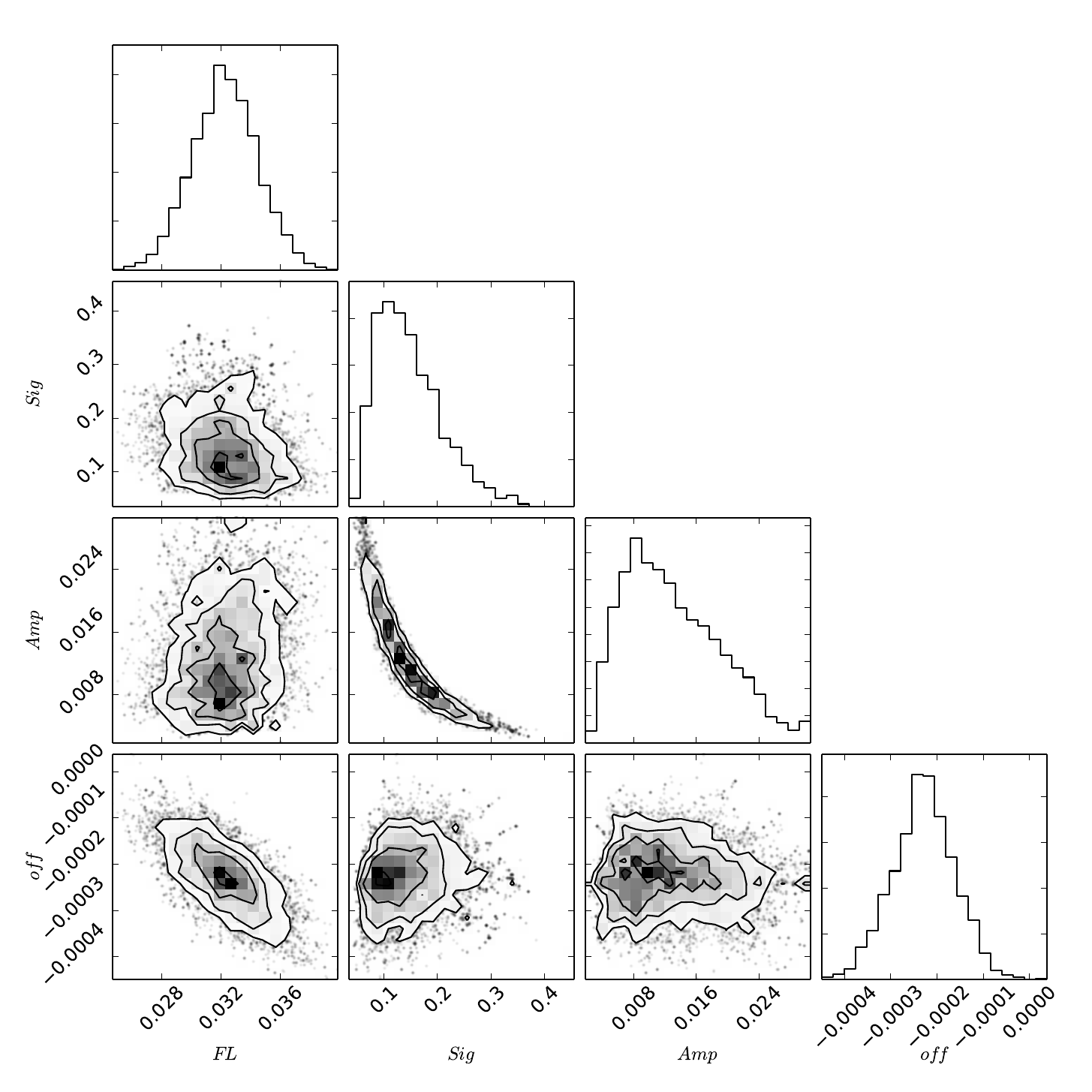}
\caption{Posterior distributions of the model parameters fitted in
  this work in the shape of histograms, along with their correlation
  plots. Values are computed from the \mbox{1.5 \AA} excess
  light curve around the sodium D$_{2}$ line. Here \textbf{FL} is the flare scaling parameter, \textbf{Sig} is width of the exoplanetary Gaussian profile, \textbf{Amp} is the amplitude of the exoplanetary Gaussian profile, and  \textbf{off} is the normalization constant for the flare model.}
\label{banana}
\end{figure*}

\end{document}